\documentclass[12pt]{article}
\usepackage{booktabs}
\usepackage{xcolor}
\usepackage{multirow}
\usepackage{comment}
\usepackage{amsmath}
\usepackage{amssymb}
\usepackage{graphicx}
\usepackage{here}
\usepackage{subcaption}
\usepackage{url}

\usepackage[sort&compress, numbers, merge]{natbib}
\usepackage{tcolorbox}
\usepackage{bbm}
\usepackage{physics}

\usepackage[compat=1.1.0]{tikz-feynhand}
\usepackage{slashed}
\usepackage{mathtools}
\usepackage{comment}
\usetikzlibrary{calc,arrows.meta,decorations.markings}

\usepackage[subpreambles=true]{standalone} 
\usepackage{todonotes}

\def\s12{s_{12}}

\usepackage[colorlinks=true, linkcolor=blue, citecolor=blue,
urlcolor=black]{hyperref}

\begin{document}

\begin{titlepage}

\vskip 1.1cm

\begin{center}

{\Large \bf 
Heavy Pseudo-Dirac Doublet Dark Matter and Non-Thermal Leptogenesis
in Light of the LZ Event
}

\vskip 1.2cm
Masahiro Ibe$^{a,b}$ and
Jun Miyamoto$^{a}$ 
\vskip 0.5cm

{\it

$^a$ {ICRR, The University of Tokyo, Kashiwa, Chiba 277-8582, Japan}

$^b$ {Kavli Institute for the Physics and Mathematics of the Universe
(WPI), \\The University of Tokyo Institutes for Advanced Study, \\ The
University of Tokyo, Kashiwa 277-8583, Japan}
}

\vskip 1.0cm

\begin{abstract}

Motivated by the recent LZ candidate nuclear-recoil event,
we study pseudo-Dirac electroweak doublet dark matter in the
mass range $10^2$--$10^4\,\mathrm{TeV}$, where bounds from
searches for neutrinos produced by dark matter annihilation
in the Sun can be evaded.
To suppress the excessive thermal relic abundance,
we consider freeze-out during inflaton domination followed
by entropy production during reheating.
We examine whether this cosmological history can also
accommodate non-thermal leptogenesis.
Our analysis includes dark matter production through the
emission of virtual electroweak gauge bosons in right-handed-neutrino
decays and through thermalization cascades.
Our estimates suggest that the two scenarios can be reconciled
if dark matter pair emission in right-handed-neutrino decays
is kinematically forbidden and cascade production is
sufficiently suppressed.
Such a realization, however, 
requires enhanced CP violation, typically
involving fine-tuning in the seesaw mechanism or the
right-handed-neutrino mass spectrum.
\end{abstract}

\end{center}
\end{titlepage}

\clearpage

\section{Introduction}
\label{sec:introduction}

The LUX-ZEPLIN (LZ) collaboration has reported a candidate
nuclear-recoil event with an energy of
$248\pm23\,\mathrm{(stat)}\pm23\,\mathrm{(sys)}\,\mathrm{keV}$
in an extended nuclear-recoil energy window~\cite{LZ:2026axp}.
Although its global significance is only $2.6\sigma$,
the event has renewed interest in inelastic dark matter,
for which endothermic scattering can suppress low-energy
recoils and favor relatively large nuclear recoil
energies~\cite{Tucker-Smith:2001myb}.
A particularly economical realization is a pseudo-Dirac
electroweak doublet fermion with a small mass splitting between its
two neutral states (see e.g. Ref.\,\cite{Bottaro:2022one}  for a systematic study of this class of models).
Their off-diagonal coupling to the $Z$ boson mediates
inelastic scattering on nuclei.
This fermion sector also arises in supersymmetric models
with nearly pure Higgsinos~\cite{Nagata:2014wma}.

Studies of the LZ event have explored Higgsino and related
fermionic dark matter models and their underlying
particle-physics realizations~\cite{Fan:2026kxx,Freese:2026sga,Wu:2026nhi,Yin:2026jnn,DiMauro:2026ldr,Visinelli:2026kgt,Smirnov:2026aqk,Du:2026guj,Ahmed:2026qjg,Bisal:2026khf,Frolovsky:2026tvq,Paul:2026okh,Khan:2026zuj,Khan:2026pgu,Girmohanta:2026qzo,Roy:2026wzb,Hor:2026voa,Jeong:2026tlf}.
Related scenarios involve singlet--doublet mixing or
additional electroweak states~\cite{Chattopadhyay:2026ryw,Borah:2026zwf,Elahi:2026vlm,Lee:2026jxl,DiMauro:2026ymt,Chakraborti:2026gmu,Lian:2026bze},
as well as scalar-doublet alternatives~\cite{Nomura:2026qyq,Murayama:2026apt}.
Complementary probes include collider searches and
experiments using different detector
targets~\cite{Kotlarski:2026pep,Cheung:2026byg,Baer:2026yrt,Maier:2026qpr,Wang:2026ssg,Graham:2026huo}.
Here we study the cosmology of the minimal pseudo-Dirac
electroweak doublet, whose annihilation rate is governed by
its electroweak interactions.

In the standard thermal freeze-out scenario, a nearly pure
electroweak doublet fermion with mass
$m_D\simeq1.1\,\mathrm{TeV}$ can account for the observed
dark matter abundance~\cite{Bottaro:2022one}.
Whether this $1.1\,\mathrm{TeV}$ thermal-doublet scenario
can also explain the LZ event is under active discussion.
Its viability depends on the predicted recoil spectrum and
event rate, the assumed local dark matter velocity distribution,
and the treatment of dark matter capture and subsequent
evolution in the
Sun~\cite{Rodd:2026tyn,Chatterjee:2026scv,Capolupo:2026uyi,Bozorgnia:2026emp}.
In particular, 
searches for neutrinos from dark matter annihilation in the
Sun impose strong constraints under the halo and
solar-capture assumptions adopted in the corresponding
analyses~\cite{Pospelov:2026ewn,DiMauro:2026dqp,Bose:2026ndd,Nguyen:2026lui,Ghosh:2026txe,Bose:2026szs}.
Although the precise allowed regions depend on these
assumptions, sufficiently heavy doublets can evade the
constraints.
This motivates the regime $m_D\sim10^2$--$10^4\,\mathrm{TeV}$~\cite{Langhoff:2026ujr,Luengas:2026tlf,Freese:2026mfi}.

For masses in this range, the annihilation cross section through the electroweak interaction is too small for freeze-out
during radiation domination to yield the observed abundance.
Avoiding this overabundance requires a departure
from the standard thermal history.
Cosmological realizations involving heavy Higgsinos or
asymmetric dark matter 
evading the constraint 
from the indirect detection
have been
explored~\cite{Langhoff:2026ujr,Nagata:2026pbj,Unwin:2026qep,Pallis:2026jdc}.
One possibility is that dark matter freezes out during an
early matter-dominated era, after which entropy production
during reheating dilutes dark matter abundance~\cite{Giudice:2000ex}.
This mechanism requires a reheating temperature below the
dark matter freeze-out temperature, $T_R<T_F$.
Such low-reheating cosmologies for heavy Higgsino dark matter
have already been studied in connection with the LZ
event~\cite{Langhoff:2026ujr,Pallis:2026jdc}.

Entropy production also dilutes a preexisting baryon
asymmetry, while a low reheating temperature restricts the
thermal production of particles responsible for baryogenesis.
In particular, conventional  thermal
leptogenesis~\cite{Fukugita:1986hr} typically requires
$T_R\gtrsim10^9\,\mathrm{GeV}$~\cite{Giudice:2003jh,Buchmuller:2004nz},
far above the reheating temperatures relevant to the
heavy-doublet scenarios.
It is therefore important to determine whether a single
reheating history can account for both the dark matter
abundance and the baryon asymmetry.

In this paper, we investigate this question in a scenario
where the inflaton decays predominantly into right-handed
neutrinos.
Dark matter freezes out during the inflaton-dominated era,
and its abundance is diluted as reheating proceeds.
The decays of the nonthermally produced right-handed
neutrinos generate a lepton asymmetry, which electroweak
sphaleron processes can convert into a baryon
asymmetry~\cite{Lazarides:1990huy,Asaka:1999jb}.
We solve the coupled Boltzmann equations to determine the
thermal contribution to the dark matter abundance and the
baryon asymmetry.

A consistent account of both abundances must also include
non-thermal dark matter production.
Even in the absence of direct inflaton decays into dark
matter, virtual electroweak radiation in right-handed-neutrino
decays can produce dark matter pairs whenever kinematically
allowed.
The thermalization cascade of energetic decay products
provides an additional source~\cite{Harigaya:2014waa}.
Both processes generate dark matter after freeze-out
and lead to overproduction.
We estimate these contributions, including re-annihilation,
and assess the resulting constraints on the reheating
scenario.

The paper is organized as follows.
Section~\ref{sec:setup} introduces the pseudo-Dirac doublet
and the non-thermal leptogenesis framework.
Section~\ref{sec: Thermal Abundance} analyzes the thermal
dark matter abundance and the baryon asymmetry during
reheating.
Section~\ref{sec:nonthermal} examines non-thermal dark matter
production in right-handed-neutrino decays and thermalization
cascades.
We conclude in section~\ref{sec:conclusions}.
Technical details are collected in the appendices.

\section{Model and Cosmological Setup}
\label{sec:setup}

\subsection{Minimal pseudo-Dirac Doublet Dark Matter}
\label{sec:doublet-model}

We introduce electroweak doublet fermions,
\begin{align}
  \chi=\begin{pmatrix}\chi^+\\ \chi^0\end{pmatrix}
      \sim\qty(\mathbf{1},\mathbf{2},+\tfrac12)\ ,
  \qquad
  \bar\chi=\begin{pmatrix}\bar\chi^0\\ \bar\chi^-\end{pmatrix}
      \sim\qty(\mathbf{1},\mathbf{2},-\tfrac12)\ ,
  \label{eq:doublet-fields}
\end{align}
where the representations refer to the
$\mathrm{SU}(3)_c\times \mathrm{SU}(2)_L\times \mathrm{U}(1)_Y$ gauge charges.
All fermions are taken to be left-handed Weyl fields, and the bar on $\bar\chi$ labels an independent left-handed field.
Together, the two fields form a vector-like electroweak doublet
with hypercharge $1/2$.
We impose a $\mathbb Z_2$ symmetry under which $\chi$ and $\bar\chi$
are odd while all the Standard Model (SM) fields are even, so that the lightest neutral state is stable.

The renormalizable dark-sector Lagrangian is
\begin{align}
  \mathcal L_{\chi}
  =i\chi^\dagger\bar\sigma^\mu D_\mu\chi
   +i\bar\chi^\dagger\bar\sigma^\mu D_\mu\bar\chi
   -m_{D}\qty(\bar\chi\cdot\chi+\mathrm{h.c.})\ ,
  \label{eq:doublet-renormalizable}
\end{align}
where $D_\mu=\partial_\mu-ig\sigma^a W^a_\mu/2-ig'/2 B_\mu$,
$\bar\sigma^\mu=(1,-\boldsymbol\sigma)$, and $A \cdot B \equiv \epsilon_{ab}A^aB^b$ with $\epsilon_{12}=1$.
The gauge coupling constants of $\mathrm{SU}(2)_L$ and $\mathrm{U}(1)_Y$ are denoted by $g$ and $g'$, respectively.
We choose field phases such that $m_D>0$.

The neutral Dirac fermion can be split into two Majorana states by
the dimension-five operators~\cite{Nagata:2014wma}
\begin{align}
  \mathcal L_{\mathrm{dim-5}}
  =-\left[
    \frac{c_\chi}{2\Lambda}(H^\dagger\chi)(H^\dagger\chi)
    +\frac{c_{\bar\chi}}{2\Lambda}
       (\bar\chi \cdot H)(\bar\chi \cdot H)
    +\mathrm{h.c.}\right],
  \label{eq:doublet-splitting-operators}
\end{align}
where $H\sim(\mathbf{1},\mathbf{2},+1/2)$ is the SM Higgs doublet,
$\Lambda\gg m_D$ is the scale of additional heavy states, and
$c_\chi$ and $c_{\bar\chi}$ are dimensionless Wilson coefficients.
We retain only these higher-dimensional interactions in the minimal
setup.

After electroweak symmetry breaking,
$\langle H\rangle=(0,v)^T$ with $v\simeq174\,\mathrm{GeV}$,
the neutral mass terms become
\begin{align}
  \mathcal L_{\mathrm{mass}}^0
  &=-\frac12    \begin{pmatrix}\chi^0&\bar\chi^0\end{pmatrix}
    \mathcal M_0
    \begin{pmatrix}\chi^0\\ \bar\chi^0\end{pmatrix}
    +\mathrm{h.c.}\ ,\\
  \mathcal M_0
  &=\begin{pmatrix}\mu_\chi&m_D\\ m_D&\mu_{\bar\chi}\end{pmatrix}\ ,
  \quad
  \mu_\chi=\frac{c_\chi v^2}{\Lambda}\ ,
  \quad
  \mu_{\bar\chi}=\frac{c_{\bar\chi}v^2}{\Lambda}\ .
  \label{eq:doublet-majorana-masses}
\end{align}
We work in the pseudo-Dirac regime
$\max\qty[|\mu_\chi|,|\mu_{\bar\chi}|]\ll m_D$.
The physical neutral masses, obtained by Takagi decomposition of
$\mathcal M_0$, satisfy
\begin{align}
  &m_{\chi_1,\chi_2}
=m_D\mp\frac12|\mu_\chi^*+\mu_{\bar\chi}|\ .
  \label{eq:doublet-neutral-masses}
\end{align}
Here $\chi_1$ and $\chi_2$ denote the lighter and heavier neutral
Majorana states, respectively, and $m_{\mathrm{DM}}\equiv m_{\chi_1}$.

Assuming that the doublet accounts for all of the dark matter,
an interpretation of the LZ event consistent with solar-capture
bounds favors the parameter region~\cite{Langhoff:2026ujr,Luengas:2026tlf,Freese:2026mfi}
\begin{align}
    &m_D\simeq m_{\mathrm{DM}}
    \sim(10^2\text{--}10^4)\,\mathrm{TeV}\ ,
    \label{eq:lz-heavy-mass-range}\\
    &\delta=m_{\chi_2}-m_{\chi_1}
    \sim300\text{--}500\,\mathrm{keV}\ .
    \label{eq:lz-splitting-range}
\end{align}
A mass splitting in this range corresponds to a cutoff scale
\begin{align}
    \Lambda\sim10^{8}\,\mathrm{GeV}
    \qty(\frac{300\,\mathrm{keV}}{\delta})\ ,
\end{align}
for Wilson coefficients of order unity.
Since $m_D\ll\Lambda$, the effective description is adequate
for the low-energy properties of the doublet.

Notice that the charged components form a Dirac fermion $\chi^\pm$ with
the tree-level mass $m_D$.
Electroweak radiative corrections raise its mass relative to the
neutral pair. In the heavy-doublet limit, $m_D \gg m_Z$, the leading
charged--neutral mass difference is~\cite{Nagata:2014wma}
\begin{align}
  \delta_\pm\equiv m_{\chi^\pm}-m_{\chi_1}
  \simeq\frac{\delta}{2}+\Delta_{\mathrm{EW}}\
   ,
  \qquad
\Delta_{\mathrm{EW}}\simeq350\,\mathrm{MeV}
  \label{eq:doublet-charged-splitting} \ .
\end{align}
Thus the lighter neutral state is the dark matter candidate.

\subsection{Non-thermal Leptogenesis}

To dilute the excessive thermal relic abundance of the heavy
doublet, we consider dark matter freeze-out during an
inflaton-dominated era, followed by entropy production
during reheating.
To account for the baryon asymmetry within the same
cosmological history, we consider non-thermal
leptogenesis~\cite{Lazarides:1990huy,Asaka:1999jb}.

In our set up, the inflaton $\Phi$ decays into right-handed neutrinos
$\bar{N}_{Ri}$ ($i=1,2,3$), which are assumed to decay promptly
into the SM Higgs and lepton doublets.
The CP violation in neutrino Yukawa interactions induces lepton number asymmetry \cite{Fukugita:1986hr,Covi:1996wh,Buchmuller:1997yu} as
\begin{align}
    \epsilon_i \equiv- \frac{\Gamma(\bar{N}_{Ri} \to H + \ell) - \Gamma(\bar{N}_{Ri} \to H^\dagger + \ell^\dagger)}{\Gamma(\bar{N}_{Ri} \to H + \ell) +
    \Gamma(\bar{N}_{Ri} \to H^\dagger + \ell^\dagger)}\ ,
\end{align}
where flavor indices of lepton doublets are suppressed. For detailed expressions of $\epsilon_i$ involving neutrino Yukawa couplings and right-handed neutrino masses $M_{N_i}$, refer to Ref.\,\cite{Covi:1996wh}.

With the above CP violation,
the inflaton decay ends up with the $B-L$ asymmetry through the 
time evolution,
\begin{align}
       \frac{dn_{B-L}}{dt} + 3 H n_{B-L} &= 2 \Gamma_\Phi \epsilon \frac{\rho_\Phi}{m_\Phi}\ , 
       \label{eq: asymmetry development}
       \\
      \epsilon &=\sum_i \mathrm{Br}_{N_i} \epsilon_i 
      \label{eq: epsilon}
\end{align}
Here, $n_{B-L}$ denotes the 
number density of the $B-L$ asymmetry, $H$ the Hubble expansion rate,
$\rho_\Phi$ the energy density of the inflaton, 
$\Gamma_\Phi$ the decay rate of the inflaton.   
The branching fractions of the inflaton into three right-handed neutrinos, $\mathrm{Br}_{N_i}$, 
are non-vanishing 
only when kinematically allowed.
In the following,
we assume the
CP-violating phases in the right-handed neutrino sector such that 
$\epsilon$  is positive.

An important feature of non-thermal leptogenesis is that
lepton asymmetry production continues until reheating is
completed at $T_R$, and hence persists after dark matter
freezes out during inflaton domination.
Successful baryogenesis, however, requires electroweak
sphaleron processes to convert the lepton asymmetry into
a baryon asymmetry.
We therefore focus on reheating temperatures
$T_R\gtrsim\order{100}\,\mathrm{GeV}$, assuming that
sphaleron conversion remains sufficiently efficient.%
\footnote{
In the SM with a radiation-dominated expansion history, sphaleron processes freeze out at
$T_{\mathrm{sph}}\simeq 132\,\mathrm{GeV}$~\cite{DOnofrio:2014rug}.
}

\section{Thermal Dark Matter Abundance}
\label{sec: Thermal Abundance}
We evolve the $B-L$ asymmetry
described in the preceding subsection together with
the inflaton and radiation energy densities and the dark matter
number density:
\begin{align}
    \frac{d\rho_\Phi}{dt}+3H\rho_\Phi
    &=-\Gamma_\Phi\rho_\Phi\ ,
    \label{eq:inflaton_development}\\
    \frac{d\rho_R}{dt}+4H\rho_R
    &=\Gamma_\Phi\rho_\Phi
       \ ,
\label{eq:radiation_development}\\
    \frac{dn_D}{dt}+3Hn_D
    &=-\langle\sigma_\mathrm{eff} v\rangle
      \left(n_D^2-n_{D,\mathrm{eq}}^2\right)\ .
    \label{eq:darkmatter_development}
\end{align}
Here, $\rho_R$ denotes the energy density of radiation, while $n_D$ and $n_{D,\mathrm{eq}}$ denote the dark matter number density and its thermal equilibrium value, respectively.
Note that we neglect the backreaction of dark matter production from the radiation bath on the radiation energy density.

Equation~\eqref{eq:radiation_development} shows that the inflaton begins to inject energy into the ``radiation'' component immediately after the end of inflation, as the universe enters the inflaton-oscillation-dominated era.
However, as discussed in Ref.\,\cite{Harigaya:2013vwa}, the decay products do not thermalize instantaneously.
Before thermalization is completed, the decay products of the inflaton (through subsequent right-handed neutrino decays) consist of energetic hard particles as well as lower-energy particles generated through their successive splittings and cascade processes.
In the following, we use $\rho_R$ to denote the total energy density carried by all these particles, irrespective of whether they have reached thermal equilibrium.
For the moment, we also assume that the inflaton does not decay directly into dark matter.
We will revisit this assumption, together with the effects associated with the non-instantaneous thermalization, in the next section.

The thermally averaged effective annihilation cross section,
$\langle \sigma_\mathrm{eff} v \rangle$, used in our numerical
analysis is given in Appendix~\ref{app: cross section}.%
\footnote{
Sommerfeld effects are also discussed in
Appendix~\ref{app: cross section}.
For the heavy doublet fermion, these effects modify the
thermally averaged cross section by approximately $1\%$
for $x=m_D/T=10$--$30$.
}
In the non-relativistic limit, it reduces to
\begin{align}
  \langle\sigma_\mathrm{eff} v\rangle
    \simeq
    \frac{81g^4+43g'^4+12g^2g'^2}
         {2048\pi m_D^2}\ ,
  \label{eq:annihilation in NR}
\end{align}
see, e.g., Ref.\,\cite{Cirelli:2005uq}.
Our normalization is a factor of two smaller than that
in Ref.\,\cite{Cirelli:2005uq}, owing to the convention
adopted in the Boltzmann equation,
Eq.\,\eqref{eq:darkmatter_development}.

So far, we have used the reheating temperature $T_R$
without specifying its precise definition.
For the following analysis, we adopt the convention
\begin{align}
    T_R \equiv
    \qty(\frac{10}{\pi^2 g_*})^{1/4}
    \sqrt{\Gamma_\Phi M_{\mathrm{Pl}}}\ ,
    \label{eq:TRH_Def}
\end{align}
where $M_{\mathrm{Pl}}\simeq2.4\times10^{18}\,\mathrm{GeV}$
is the reduced Planck mass.

We solve the coupled Boltzmann equations above with
the initial conditions
\begin{align}
    \rho_\Phi = 3M_{\mathrm{Pl}}^2 H_i^2\ , \quad
    \rho_R = 0\ , \quad
    n_{B-L} = 0\ , \quad
    n_D = 0\ ,
\end{align}
where $H_i$ is the initial Hubble expansion rate.
We take the effective number of relativistic degrees
of freedom to be constant, $g_*=106.75$.

\subsection{Inflaton and Radiation}

Before proceeding to the numerical analysis, let us first examine how the
inflaton and radiation components evolve.
Well before the completion of reheating, i.e., for
$H\gg\Gamma_\Phi$, the inflaton energy density scales as
\begin{align}
    \rho_\Phi \propto a^{-3}\ ,
\end{align}
where $a$ is the scale factor of the universe.
Equation~\eqref{eq:radiation_development} then implies
\begin{align}
    \rho_R \propto a^{-3/2}\ .
    \label{eq: rhor scaling}
\end{align}
As noted above, $\rho_R$ collectively denotes the energy density carried by the
energetic inflaton decay products and the lower-energy particles, irrespective of whether they have
reached thermal equilibrium.

The thermalization of these particles during the inflaton-dominated era
proceeds through a sequence of processes, as discussed in
Refs.\,\cite{Harigaya:2013vwa}.
The inflaton decay products are initially highly energetic and far from
thermal equilibrium.
Through gauge interactions, 
they radiate lower-energy particles, which gradually populate the soft momentum modes.
As the occupation of these soft modes increases, interactions among the soft
particles become efficient, eventually establishing a soft thermal bath.

Once the soft thermal bath with temperature $T$ is established, the energy
degradation of the remaining hard particles can be characterized by their
splitting rate.
For a hard particle with momentum $k\gg T$, the splitting rate, including
the Landau--Pomeranchuk--Migdal effect, is given  by
Refs.\,\cite{Harigaya:2013vwa,Harigaya:2014waa} as
\begin{align}
    \Gamma_{\rm split}(k)
    \sim
    \alpha_2^2 T
    \sqrt{\frac{T}{k}}\, ,
    \label{eq:splitting_rate}
\end{align}
where we take the $\mathrm{SU}(2)_L$ gauge interaction to be responsible
for the thermalization of the hard leptons and Higgs bosons produced through
the right-handed neutrino decays.
Since the splitting rate decreases with momentum as
$\Gamma_{\rm split}(k)\propto k^{-1/2}$, while the energy transferred in
each splitting can be of order $k$, the energy degradation of a hard particle
proceeds efficiently through splittings into daughter particles with momenta
of $\mathcal{O}(k)$.
The energy initially stored in the hard particles, $E_i$, 
is therefore transferred
toward lower energies through a splitting cascade, which may be illustrated
schematically as
\begin{align}
    E_i
    \;\rightarrow\;
    2\times{\cal O}(E_i/2)
    \;\rightarrow\;
    4\times{\cal O}(E_i/4)
    \;\rightarrow\;\cdots\, .
\end{align}

The largest momentum that can be efficiently degraded within one Hubble
time is determined by
$\Gamma_{\rm split}(k_{\rm split})\sim H$, yielding
\begin{align}
    k_{\rm split}
    \sim
    \frac{\alpha_2^4 T^3}{H^2}\, .
    \label{eq:ksplit}
\end{align}
As the thermal bath develops, $k_{\rm split}$ increases with time.
Once $k_{\rm split}$ reaches $\mathcal{O}(m_\Phi)$, the energy injected by inflaton decays can be regarded as being thermalized promptly.

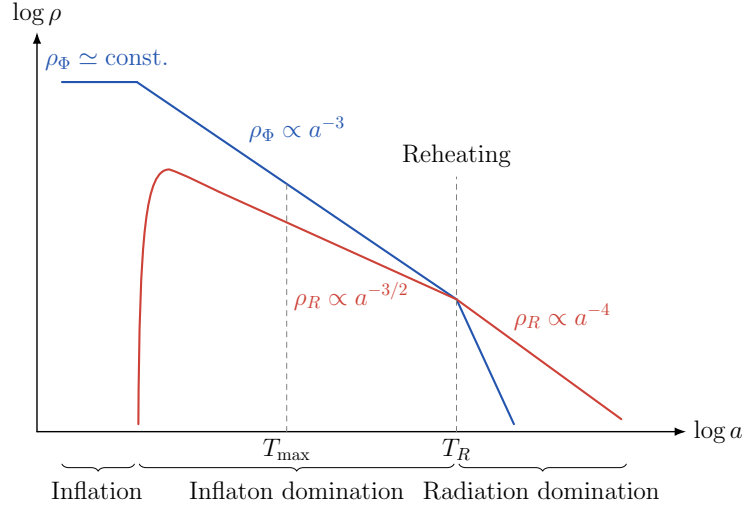
\begin{figure}
    \centering
\begin{tikzpicture}[
    >=Latex,
    font=\small,
    line cap=round,
    line join=round
]


\def\xmax{13}
\def\ymax{8}

\def\xendinflation{2.0}
\def\xthermal{5.0}
\def\xreheat{8.4}

\definecolor{inflatoncolor}{RGB}{35,85,175}
\definecolor{radiationcolor}{RGB}{205,65,55}


\draw[->, thick]
    (0,0) -- (\xmax,0)
    node[right,font=\large] {$\log a$};

\draw[->, thick,font=\large]
    (0,0) -- (0,\ymax)
    node[above] {$\log \rho$};


\draw[
    inflatoncolor,
    very thick
]
    (0.5,7.0)
    -- (2.0,7.0);

\draw[
    inflatoncolor,
    very thick
]
    (2.0,7.0)
    -- (8.4,2.65);

\draw[
    inflatoncolor,
    very thick
]
    (8.4,2.65)
    -- (9.55,0.15);


\draw[
    radiationcolor,
    very thick
]
 (2.03,0.15)
    .. controls (2.05,4)
                and (2.18,5.25)
    .. (2.65,5.25)
    .. controls (2.95,5.15)
                and (3.35,4.94)
    .. (3.65,4.80);

\draw[
    radiationcolor,
    very thick
]
    (3.65,4.80)
    -- (8.4,2.65);

\draw[
    radiationcolor,
    very thick
]
    (8.4,2.65)
    -- (11.7,0.25);


\node[
    inflatoncolor,
    anchor=south,
    font=\large
] at (1.40,7.10)
    {$\rho_\Phi\simeq\mathrm{const.}$};

\node[
    inflatoncolor,
    anchor=south west,
    font=\large
] at (4.1,5.65)
    {$\rho_\Phi\propto a^{-3}$};

\node[
    radiationcolor,
    anchor=south west,
    font=\large
] at (5.,2.30)
    {$\rho_R\propto a^{-3/2}$};

\node[
    radiationcolor,
    anchor=south west,
    font=\large
] at (9.40,1.90)
    {$\rho_R\propto a^{-4}$};





\draw[
    gray,
    dashed,
    thin
]
    (\xthermal,0)
    -- (\xthermal,5.0);

\node[
    below,
    font=\large
] at (\xthermal,0)
    {$T_{\max}$};


\draw[
    gray,
    dashed,
    thin
]
    (\xreheat,0)
    -- (\xreheat,5.1);

\node[
    align=center,
    anchor=south,
    font=\large
] at (\xreheat,5.15)
    {Reheating};

\node[
    below,
    font=\large
] at (\xreheat,0)
    {$T_R$};


\draw[
    decorate,
    decoration={
        brace,
        mirror,
        amplitude=4pt
    }
]
    (0.5,-0.65) -- (1.95,-0.65);

\node[
    below,
    align=center,
    font=\large
] at (1.2,-0.85)
    {Inflation};

\draw[
    decorate,
    decoration={
        brace,
        mirror,
        amplitude=4pt
    }
]
    (2.05,-0.65) -- (8.35,-0.65);

\node[
    below,
    font=\large
] at (5.2,-0.85)
    {Inflaton domination};

\draw[
    decorate,
    decoration={
        brace,
        mirror,
        amplitude=4pt
    }
]
    (8.45,-0.65) -- (11.8,-0.65);

\node[
    below,
    font=\large
] at (10.1,-0.85)
    {Radiation domination};

\end{tikzpicture}
    \caption{Schematic evolution of the inflaton and radiation energy densities during and after inflation. During inflation, the inflaton energy density remains approximately constant. 
    After inflation, the Universe enters an inflaton-dominated era, in which $\rho_\Phi\propto a^{-3}$, while radiation produced by inflaton decays evolves as $\rho_R\propto a^{-3/2}$ after its initial build-up. 
    By the time that
    the thermal bath with $T_\mathrm{max}$ is achieved, $\rho_R$ collectively denotes the energy density carried by the
   energetic inflaton decay products and the lower-energy particles generated
through their subsequent splittings.    
At $T_{\max}$, the thermalization of the decay products becomes efficient. The reheating process is completed around $T=T_R$, after which the universe becomes radiation dominated and $\rho_R\propto a^{-4}$.}
    \label{fig:placeholder}
\end{figure}

Around this epoch, the energy density of the thermal bath is given by,
\begin{align}
    \rho_R \sim \frac{\pi^2 g_*}{30}T^4 
    \sim
    \frac{\Gamma_\Phi}{H}\rho_\Phi
    \sim
    3\Gamma_\Phi M_{\rm Pl}^2 H\, .
    \label{eq:T_H_thermalization}
\end{align}
Combining this relation with Eq.\,\eqref{eq:ksplit} and
$k_{\rm split}\sim m_\Phi$, we obtain the maximum temperature of the
thermal bath,
\begin{align}
    T_{\rm max}
    &\sim
    \alpha_2^{4/5}
    \qty(\frac{90}{\pi^2 g_*})^{2/5}
    \left(
        \frac{\Gamma_\Phi M_{\rm Pl}^2}{m_\Phi^3}
    \right)^{2/5}
    m_\Phi\, ,\\
    &\sim 4\times 10^6\,\mathrm{GeV}
    \qty(
\frac{T_R}{10^{3}\,\mathrm{GeV}}    )^{4/5} 
\qty(\frac{10^{10}\,\mathrm{GeV}}{m_\Phi})^\frac{1}{5}
\label{eq:Tmax_thermalization}\ ,
\end{align}
which satisfies $T_{\rm max}\gg T_R$.
After the temperature
of the radiation energy density
reaches $T_{\rm max}$, the energy injected into the
radiation sector is efficiently thermalized, and $\rho_R$ can  be
identified with the usual thermal radiation energy density.

Once the radiation component $\rho_R$ has thermalized, its temperature scales as
\begin{align}
T \propto a^{-3/8}\ ,
\label{eq: T scaling}
\end{align}
for $T_R \lesssim T \lesssim T_\mathrm{max}$, as follows from the scaling of the radiation energy density in Eq.\,\eqref{eq: rhor scaling}.
In this scaling period, the energy densities can be parameterized by,
\begin{align}
    &\rho_\Phi(T) \simeq  
    \qty(\frac{1-r_R}{r_R})\times\rho_R(T_R)
    \qty(\frac{T}{T_R})^{8}\ ,
    \label{eq: rhoPhi in ID}
    \\
   &\rho_R(T) \simeq \rho_R(T_R)\
    \qty(\frac{T}{T_R})^4\ .
    \label{eq: rhoR in ID}
\end{align}
Here, $r_R$ denotes the fraction of the total energy density carried by radiation at $T=T_R$,
\begin{align}
r_R \equiv
\frac{\rho_R}{\rho_{\mathrm{tot}}}
\bigg|_{T=T_R}
\simeq 0.6\ ,
\label{eq: rr}
\end{align}
where $T_R$ is defined in Eq.\,\eqref{eq:TRH_Def}. The numerical value in Eq.\,\eqref{eq: rr} is obtained by solving Eqs.\,\eqref{eq:inflaton_development} and \eqref{eq:radiation_development}.
The value of $r_R$
can also be analytically estimated as discussed in the Appendix~\ref{app: Energy Density Ratio}

After the completion of the reheating, i.e., 
$H\ll \Gamma_\Phi$,
the radiation energy density dominates.
Accordingly, the energy density
and the radiation temperature scale by
\begin{align}
    \rho_R(T) \propto a^{-4}  \ , \qquad T\propto a^{-1} \ ,
\end{align}
for $T \ll T_R$.
The inflaton energy density disappears exponentially.

Note that we allow scenarios in which the inflaton mass $m_\Phi$ lies
well above the cutoff, i.e.,  $m_\Phi\gg\Lambda$.
The inflaton decay chain may then involve additional states
associated with the UV completion of the operators in
Eq.\,\eqref{eq:doublet-splitting-operators}.
We assume that these states decay promptly into doublet fermions and SM particles, so that additional dark matter abundance to be negligible.

\subsection{Baryon Asymmetry}
We next estimate the baryon asymmetry generated through
non-thermal leptogenesis.
The dominant contribution to the final asymmetry arises
near the end of reheating, giving
\begin{align}
    Y_{B-L}\equiv \frac{n_{B-L}}{s_R}\bigg|_{T_R}
    \sim 2\epsilon\,
    \frac{n_\Phi}{s_R}\bigg|_{T_R}\ .
\end{align}
Here, $\epsilon$ is the effective lepton asymmetry defined in
Eq.\,\eqref{eq: epsilon}, and the factor of two accounts
for the two right-handed neutrinos produced in each
inflaton decay.
The entropy density is denoted by $s_R$.
We assume that the relevant right-handed neutrinos satisfy
$M_{N_i}\gg T_R$ and that washout of the lepton asymmetry is negligible.

At $T=T_R$, the inflaton number density $n_\Phi$ satisfies
\begin{align}
    \frac{n_\Phi}{s_R}\bigg|_{T_R}
    &=
    \frac{\rho_\Phi}{\rho_R}
    \frac{\rho_R}{m_\Phi s_R}\bigg|_{T_R}
    =
    \qty(\frac{1-r_R}{r_R})\times
    \frac{3}{4}\frac{T_R}{m_\Phi}
    \simeq 0.5\,\frac{T_R}{m_\Phi}\ ,
\end{align}
where 
we have used $\rho_R/s_R|_{T_R}=3T_R/4$.
The final $B-L$ yield is therefore approximately
\begin{align}
    Y_{B-L}
    \sim \epsilon\frac{T_R}{m_\Phi}\ .
    \label{eq:B-L_yield}
\end{align}
This expression provides an order-of-magnitude estimate.
A more precise result follows from
solving Eq.\,\eqref{eq: asymmetry development} together
with the equations for the inflaton and radiation
energy densities.

Assuming that asymmetry production is completed while electroweak
sphaleron processes remain in equilibrium, the resulting baryon
yield is given by
\begin{align}
    Y_B
    =c_{\mathrm{sph}}Y_{B-L}
    \sim c_{\mathrm{sph}}\epsilon
    \frac{T_R}{m_\Phi}\ ,
    \label{eq: YB}
\end{align}
where $c_{\mathrm{sph}}=28/79$ is the sphaleron conversion
factor~\cite{Khlebnikov:1988sr,Harvey:1990qw}.
Thus, for fixed $\epsilon$ and $m_\Phi$, the baryon
asymmetry scales linearly with $T_R$.
As we will see below, reproducing the thermal dark matter
abundance in the heavy-mass regime of interest requires
reheating temperatures of order a TeV.
Such low reheating temperatures consequently require
a relatively large CP asymmetry to account for the
observed baryon abundance.

\subsection{Dark Matter Freeze-Out During Reheating}

We now discuss the qualitative behavior of the thermal
dark matter abundance.
We assume that dark matter reaches chemical equilibrium
with the SM thermal bath during inflaton domination
and freezes out well before reheating is completed.
The freeze-out temperature $T_F$ therefore satisfies
$T_R\ll T_F<T_{\mathrm{max}}$.
The condition for chemical equilibrium will be discussed below.

During inflaton domination, when $H\gg\Gamma_\Phi$,
the Hubble expansion rate scales approximately as
\begin{align}
    H\simeq H_R\left(\frac{x}{x_R}\right)^{-4}\ ,
    \qquad
    x\equiv\frac{m_D}{T}\ ,
    \qquad
    x_R\equiv\frac{m_D}{T_R}\ ,
    \label{eq: H in inflaton domination}
\end{align}
where $H_R$ is the Hubble expansion rate at $T=T_R$
(see Eqs.\,\eqref{eq: rhoPhi in ID}
and \eqref{eq: rhoR in ID}).
In this regime, the Boltzmann equation,
Eq.\,\eqref{eq:darkmatter_development}, becomes
\begin{align}
    \frac{dn_D}{dx}+\frac{8}{x}n_D
    =-\frac{8\langle\sigma_\mathrm{eff} v\rangle}{3Hx}
    \left(n_D^2-n_{D,\mathrm{eq}}^2\right)\ ,
    \label{eq:Boltzmann-Eq-x}
\end{align}

After freeze-out at $x_F\equiv m_D/T_F$,
the equilibrium contribution becomes negligible, giving
\begin{align}
    \frac{d}{dx}\left(\frac{1}{x^8n_D}\right)
    \simeq
    \frac{8\langle\sigma_\mathrm{eff} v\rangle}{3H_Rx_R^4}\,x^{-5}\ .
\end{align}
Approximating $\langle\sigma_\mathrm{eff} v\rangle$ by its
non-relativistic limit in Eq.\,\eqref{eq:annihilation in NR}
and integrating from $x_F$ to $x_R\gg x_F$, we obtain
\begin{align}
    \frac{1}{n_D(x_R)}
    \simeq
    \frac{1}{n_D(x_F)}
    \left(\frac{x_R}{x_F}\right)^8
    +\frac{2\langle\sigma_\mathrm{eff} v\rangle}{3H_R}
    \left(\frac{x_R}{x_F}\right)^4\ .
    \label{eq: nDxR}
\end{align}
The first term describes the dilution of the abundance
present at freeze-out by cosmic expansion, while the
second accounts for continued annihilation between
freeze-out and reheating.

At freeze-out, $n_D(x_F)$ is of order
$n_{D,\mathrm{eq}}(x_F)$, with
\begin{align}
    &n_{D,\mathrm{eq}}(x_F)
    \sim
    \frac{H_F}{\langle\sigma_\mathrm{eff} v\rangle}x_F
    =
    \frac{H_R}{\langle\sigma_\mathrm{eff} v\rangle}
    \left(\frac{x_R}{x_F}\right)^4x_F\ ,\\
    &n_{D,\mathrm{eq}}(x_F)
    =g_D m_D^3
    \qty(2\pi x_F)^{-3/2}e^{-x_F} \ ,
\end{align}
where $g_D = 8$ and 
$H_F\equiv H(T_F)$
(see Eq.\,\eqref{eq:at xF}).
For the parameter range of interest, this condition gives
$x_F=\order{10}$.
The first term on the right-hand side of
Eq.\,\eqref{eq: nDxR} is therefore suppressed relative
to the second by a factor of order $x_F^{-1}$.
Keeping the dominant second term, we estimate the
dark matter yield at $T=T_R$ as
\begin{align}
    Y_D
    \equiv\frac{n_D(x_R)}{s_R}
    \sim
    \frac{3H_R}{2s_R\langle\sigma_\mathrm{eff} v\rangle}
    \left(\frac{x_F}{x_R}\right)^4\ .
    \label{eq: YDapp}
\end{align}

Substituting $H_R = \sqrt{\rho_R/(3M_{\rm{Pl}}^2 r_R)}$ into Eq.~\eqref{eq: YDapp}, and using 
$x_F/x_R=T_R/T_F$, we obtain
\begin{align}
    Y_D
    \sim
    \sqrt{\frac{90}{\pi^2g_* r_R}}
    \frac{3}{8\langle\sigma_\mathrm{eff} v\rangle
    M_{\mathrm{Pl}}T_F}
    \left(\frac{T_R}{T_F}\right)^3\ .
    \label{eq: YD}
\end{align}
The observed dark matter abundance,
$\Omega_{\mathrm{DM}}h^2\simeq0.12$~\cite{Planck:2018vyg},
corresponds to
\begin{align}
    m_DY_D
    \simeq4.4\times10^{-10}\,\mathrm{GeV}
    \qty(\frac{\Omega_{\mathrm{DM}}h^2}{0.12})\ .
    \label{eq: upper bound on YD}
\end{align}
Matching Eq.\,\eqref{eq: YD} to this value determines
the reheating temperature required to account for the observed abundance.
We denote this temperature by $T_R^{(\mathrm{th})}$ and find
\begin{align}
    T_R^{(\mathrm{th})}
    \simeq1.3\times10^4\,\mathrm{GeV}
    \qty(\frac{m_D}{10^2\,\mathrm{TeV}})^{1/3}
    x_F^{-4/3}\ .
    \label{eq: required TR}
\end{align}

We next examine the mass dependence of $T_R^{(\mathrm{th})}$.
In addition to the explicit $m_D^{1/3}$ dependence in
Eq.\,\eqref{eq: required TR}, there is an implicit dependence
through $x_F$.
Along the relation that reproduces the observed abundance,
the mass dependence of $x_F$ is given by
\begin{align}
    \dv{\ln x_F}{\ln m_D}
    \simeq-\frac{7}{3}
    \qty(
        x_F+\frac{1}{6}+\frac{2x_F}{2x_F-13}
    )^{-1}\ ,
\end{align}
as derived in Appendix~\ref{app: freeze-out temperature}.
The required reheating temperature therefore scales
approximately as
\begin{align}
    T_R^{(\mathrm{th})}
    &\propto m_D^{\gamma_R}\ ,
    \qquad
    \gamma_R
    \equiv\dv{\ln T_R^{(\mathrm{th})}}{\ln m_D}
    \simeq\frac{1}{3}
    -\frac{4}{3}\dv{\ln x_F}{\ln m_D}
    \sim0.4\text{--}0.6\ ,
    \label{eq: gammaR}
\end{align}
for $x_F=\order{10}$.
Thus, $T_R^{(\mathrm{th})}$ is of order a TeV over
the mass range
$10^2\,\mathrm{TeV}\lesssim m_D\lesssim10^4\,\mathrm{TeV}$.

Before closing this subsection, we examine whether
dark matter reaches chemical equilibrium with the
SM thermal bath during inflaton domination.
Assuming $m_D<T_{\mathrm{max}}$, an approximate criterion is
\begin{align}
    \max_{T_R\ll T<T_{\mathrm{max}}}
    \qty[        \frac{n_{D,\mathrm{eq}}        \expval{\sigma_{\mathrm{eff}}v}}{H}
    ]
    \gtrsim1\ .
    \label{eq: chemical equilibrium}
\end{align}
Using the thermally averaged cross section in
Eq.\,\eqref{eq:thermal-irrep-unexpanded}, we find numerically
that the ratio inside the brackets is maximized at
\begin{align}
    x_*\equiv\frac{m_D}{T_*}\simeq3\ .
\end{align}
During inflaton domination, the ratio 
$n_{D,\mathrm{eq}}\expval{\sigma_{\mathrm{eff}}v}/H$
is proportional
to $T_R^2$ at fixed temperature $T$, as follows from
Eq.\,\eqref{eq: H in inflaton domination}.
The equilibrium criterion can therefore be expressed as
\begin{align}    T_R\gtrsim1\,\mathrm{GeV}
    \qty(\frac{m_D}{10^2\,\mathrm{TeV}})^{3/2}\ .
    \label{eq: TRH_Bound}
\end{align}
This estimate includes relativistic corrections to the
thermal average.
At $T_*\simeq m_D/3$, the thermally averaged cross section
is approximately $0.4$ times its non-relativistic limit
in Eq.\,\eqref{eq:annihilation in NR}.
For reheating temperatures of order a TeV, the criterion
is therefore satisfied for dark matter masses up to
approximately $10^4\,\mathrm{TeV}$, provided that
$T_R\ll T_*\simeq m_D/3<T_{\mathrm{max}}$.

\subsection{Numerical Analysis}
\label{sec:Numerical Analysis}

We numerically solve
Eqs.\,\eqref{eq:inflaton_development}--\eqref{eq:darkmatter_development}
to track dark matter production during reheating.
As discussed above, the thermal bath is established at
$T=T_{\max}$, with $T_{\max}$ given by
Eq.\,\eqref{eq:Tmax_thermalization}, at a time later than
the initial epoch $H=H_i$.
Since we focus on dark matter production from the thermal bath,
we begin evolving the Boltzmann equation for the thermal dark
matter abundance at a chosen initial temperature
$T_{\max}^{(\mathrm{init})}$, with $n_D=0$.
We treat $T_{\max}^{(\mathrm{init})}$ as a free parameter.

This prescription neglects any dark matter produced before
the thermal bath is established.
For $T^{(\mathrm{init})}_{\max}\gtrsim m_D/5$, however, our numerical results
indicate that dark matter reaches chemical equilibrium
over the mass range of interest,
$10^2\,\mathrm{TeV}\lesssim m_D\lesssim10^4\,\mathrm{TeV}$.
Its abundance therefore loses memory of the initial
condition, making the resulting thermal relic abundance
insensitive to earlier production.

\begin{figure}[tbp]
    \centering
  \includegraphics[width=0.6\linewidth]{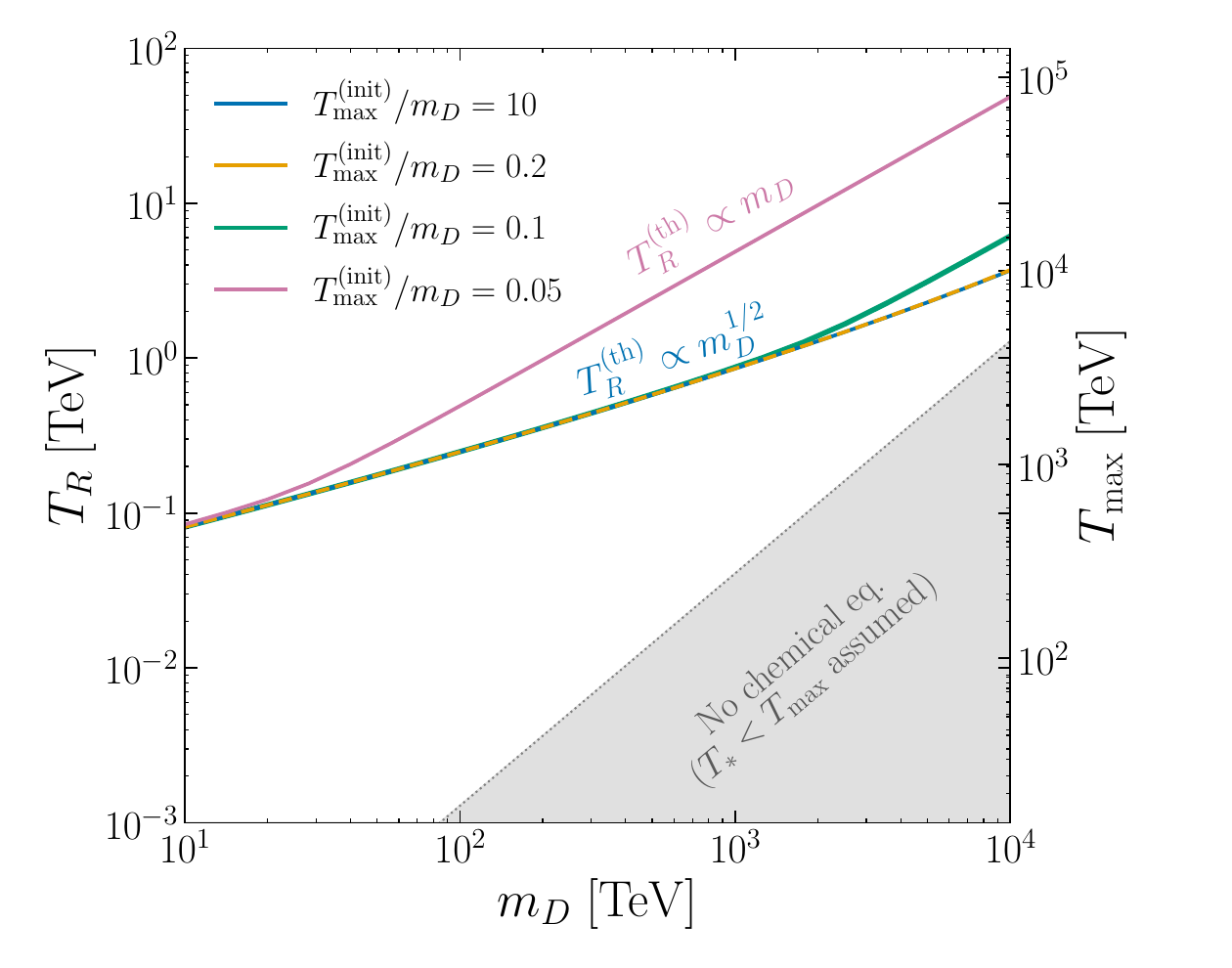}
    \caption{
    Reheating temperature $T_R^{(\mathrm{th})}$ required
    for thermal production to reproduce the observed
    dark matter abundance, shown as a function of $m_D$
    for fixed values of $T_{\max}^{(\rm{init})}/m_D$.
    The gray region does not satisfy the approximate
    chemical-equilibrium criterion in
    Eq.\,\eqref{eq: TRH_Bound}, which assumes
    $T_*<T_{\max}$.
    The right vertical axis gives a reference conversion
    from $T_R$ to $T_{\max}$ for
    $m_\Phi=10^{10}\,\mathrm{GeV}$ in Eq.\,\eqref{eq:Tmax_thermalization}.
    }
    \label{fig:TR}
\end{figure}

Figure~\ref{fig:TR} shows $T_R^{(\mathrm{th})}$ as a function
of the dark matter mass for several fixed values of
$T_{\max}^{(\rm{init})}/m_D$, requiring
$\Omega_Dh^2=0.12$~\cite{Planck:2018vyg}.
For reference, the right vertical axis shows the
corresponding $T_{\max}$ obtained from
Eq.\,\eqref{eq:Tmax_thermalization} with
$m_\Phi=10^{10}\,\mathrm{GeV}$.
The gray region indicates where the approximate
equilibrium criterion in Eq.\,\eqref{eq: TRH_Bound}
is not satisfied.
This criterion applies only when $T_*<T_{\max}$.

For $T_{\max}^{(\rm{init})}\gtrsim m_D/5$, the numerical results exhibit
the approximate scaling
$T_R^{(\mathrm{th})}\propto m_D^{1/2}$,
corresponding to $\gamma_R\simeq1/2$ in
Eq.\,\eqref{eq: gammaR}.
The required reheating temperature is nearly unchanged
between $T_{\max}^{(\rm{init})}\simeq m_D/5$ and $T_{\max}^{(\rm{init})}\gg m_D$.
Throughout the mass range
$10^2\,\mathrm{TeV}\lesssim m_D\lesssim10^4\,\mathrm{TeV}$,
the required reheating temperature is of order a TeV,
in agreement with Eq.\,\eqref{eq: required TR}
for $x_F = \order{10}$.

For comparison, we also show the case
$T_{\max}^{(\rm{init})}\simeq m_D/20$, in which thermal production
is strongly suppressed.
In this regime, the abundance depends primarily on
$T_R/m_D$, leading to the approximate scaling
$T_R^{(\mathrm{th})}\propto m_D$~\cite{Feldstein:2013uha}.

\begin{figure}[tbp]
    \centering
    \includegraphics[width=0.6\linewidth]{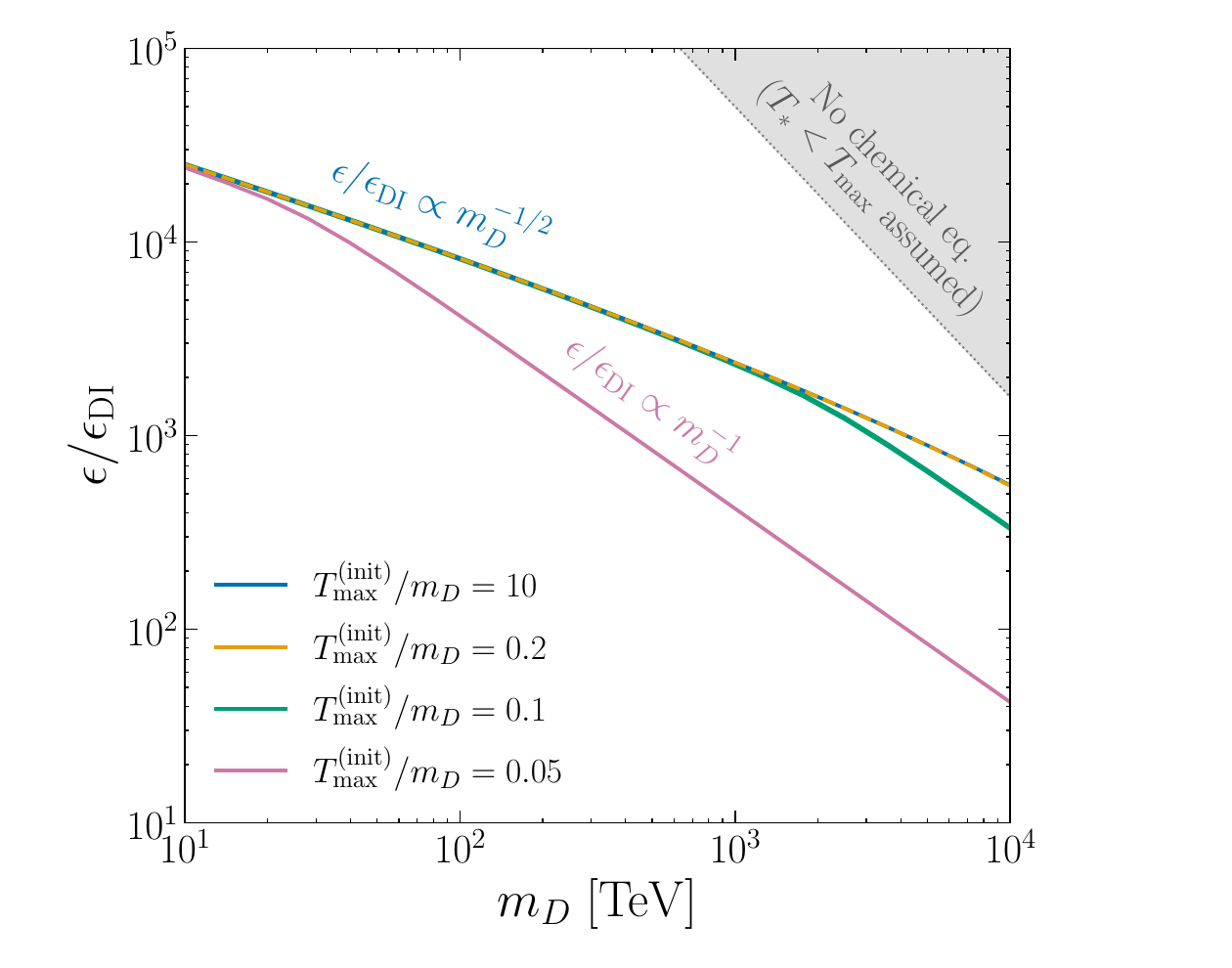}
    \caption{
    Effective CP asymmetry required to reproduce the observed
    baryon abundance, normalized to the Davidson--Ibarra bound.
    The curves correspond to fixed values of $T_{\max}^{(\rm{init})}/m_D$,
    with $T_R=T_R^{(\mathrm{th})}$ taken from
    Fig.\,\ref{fig:TR}.
    We set $m_\Phi\simeq2M_{N_1}$ to maximize the
    kinematic factor $2M_{N_1}/m_\Phi$.
    The gray region does not satisfy the approximate
    chemical-equilibrium criterion for dark matter,
    which assumes $T_*<T_{\max}$.
    }
    \label{fig:Epsilon}
\end{figure}

Figure~\ref{fig:Epsilon} shows the effective CP asymmetry
required to reproduce the observed baryon
abundance~\cite{Planck:2018vyg},
\begin{align}
    Y_B\simeq8.7\times10^{-11}\ ,
\end{align}
with the reheating temperature set to $T_R=T_R^{(\mathrm{th})}$.
In the figure, we assume that 
the inflaton predominantly decays into the lightest right-handed neutrino $N_1$ and normalize the required asymmetry to the Davidson--Ibarra
bound, 
$\epsilon_1\lesssim\epsilon_{\mathrm{DI}}$
\cite{Davidson:2002qv}, where
\begin{align}
    \epsilon_{\mathrm{DI}}
    \simeq
    \frac{3}{16\pi}
    \frac{M_{N_1}}{v^2}
    m_{\nu_3}\ .
    \label{eq:DI_bound}
\end{align}
This expression assumes a hierarchical right-handed-neutrino
spectrum, $M_{N_1}\ll M_{N_2},M_{N_3}$.
We also assume the normal ordering
of the active neutrino masses
with 
$m_{\nu_1}< m_{\nu_2} <m_{\nu_3}$.

Using this normalization, Eq.\,\eqref{eq: YB} gives
\begin{align}
    Y_B
    \sim2\times10^{-14}
    \frac{\epsilon}{\epsilon_{\mathrm{DI}}}
    \qty(\frac{T_R}{1\,\mathrm{TeV}})
    \qty(\frac{m_{\nu_3}}{0.05\,\mathrm{eV}})
    \frac{2M_{N_1}}{m_\Phi}\ .
\end{align}
For fixed $T_R$ and $\epsilon/\epsilon_{\mathrm{DI}}$,
this estimate is maximized when the inflaton mass lies
just above the right-handed-neutrino pair-production
threshold, $m_\Phi\simeq2M_{N_1}$.

From the figure, we find that the required CP asymmetry
substantially exceeds the Davidson--Ibarra bound.
Allowing for a general inflaton mass, the required
enhancement can be expressed as
\begin{align}
    \mathcal{E}
    \equiv
    \frac{\epsilon}{\epsilon_{\mathrm{DI}}}
    \frac{2M_{N_1}}{m_\Phi}
    \sim10^3\text{--}10^4\ ,
\end{align}
for $10^2\,\mathrm{TeV}\lesssim m_D\lesssim10^4\,\mathrm{TeV}$.
For fixed $m_{\nu_3}$, the required enhancement scales as
$\mathcal{E}\propto1/T_R^{(\mathrm{th})}$.
The results in Fig.\,\ref{fig:TR} therefore imply
$\mathcal{E}\propto m_D^{-1/2}$ for $T_{\max}\gtrsim m_D/5$
and $\mathcal{E}\propto m_D^{-1}$ for $T_{\max}=m_D/20$.

In the conventional hierarchical, $N_1$-dominated scenario,
the Davidson--Ibarra bound and the kinematic condition
$m_\Phi>2M_{N_1}$ imply $\mathcal{E}\lesssim1$.
Therefore, the reheating temperatures required by the thermal dark
matter abundance are  too low to account for
the observed baryon asymmetry in this scenario,
even when $m_\Phi \simeq 2 M_{N_1}$.

The above conclusion applies to an $N_1$-dominated scenario
in which the hierarchical Davidson--Ibarra bound is valid.
It does not directly extend to scenarios in which the
asymmetry is generated predominantly by decays of the
heavier state $N_2$, whose CP asymmetry is not subject
to the same bound on $\epsilon_1$~\cite{DiBari:2005st}.
Non-thermal production of heavier right-handed neutrinos
provides a possible realization~\cite{Asaka:2002zu}.

To demonstrate the possible enhanced CP asymmetry, we consider a scenario where inflaton decays predominantly into $N_2$ pairs,
with $\mathrm{Br}_{N_2}\simeq1$.
We take $m_\Phi\simeq 2M_{N_2}$,
slightly above the pair-production threshold.
In this case, the required enhancement becomes
\begin{align}
    \mathcal{E}
    =\frac{\epsilon_2}{\epsilon_{\mathrm{DI}}}
    \frac{2M_{N_1}}{m_\Phi}
    \simeq
    \frac{\epsilon_2}{\epsilon_{\mathrm{DI}}}
    \frac{M_{N_1}}{M_{N_2}}
    \ .
\end{align}
Table~\ref{tab:eps2_benchmarks} presents two benchmark points
with $\mathcal{E}\simeq10^3$ and $10^4$.
Here, we work in the flavor basis in which the charged-lepton
and right-handed-neutrino mass matrices are diagonal
and adopt the Casas--Ibarra parametrization of the
neutrino Yukawa matrix~\cite{Casas:2001sr},
\begin{gather}
    Y=\frac{1}{v}\sqrt{D_M}R\sqrt{D_m}\,
    U_{\mathrm{PMNS}}^\dagger\ ,
    \qquad
    R^TR=\mathbbm{1}\ ,
    \label{eq:CI_benchmarks}\\
    D_M=\operatorname{diag}(M_{N_1},M_{N_2},M_{N_3})\ ,
    \qquad
    D_m=\operatorname{diag}(m_{\nu_1},m_{\nu_2},m_{\nu_3})\ .
\end{gather}
The first index of $Y_{i\alpha}$ labels the heavy neutrino,
and the second labels the lepton flavor.
We parametrize the complex orthogonal matrix $R$ as
\begin{align}
    R(\theta,\phi,\psi)=
    \begin{pmatrix}
        c_\theta c_\phi &
        c_\theta s_\phi s_\psi-s_\theta c_\psi &
        c_\theta s_\phi c_\psi+s_\theta s_\psi\\
        s_\theta c_\phi &
        s_\theta s_\phi s_\psi+c_\theta c_\psi &
        s_\theta s_\phi c_\psi-c_\theta s_\psi\\
        -s_\phi & c_\phi s_\psi & c_\phi c_\psi
    \end{pmatrix}\ ,
    \label{eq:R_benchmarks}
\end{align}
where $\theta,\phi,\psi\in\mathbb{C}$,
$c_z\equiv\cos z$, and $s_z\equiv\sin z$.
We assume normal ordering of the active neutrino masses with
$(m_{\nu_1},m_{\nu_2},m_{\nu_3})
=(0,0.0087,0.050)\,\mathrm{eV}$ and use $v=174\,\mathrm{GeV}$.
The PMNS mixing angles and Dirac CP phase are fixed
to their central values in the NuFIT~6.1 analysis
(IC24 with SK-atm)~\cite{Esteban:2024eli},
and the Majorana phases are set to zero.

For both benchmarks, the largest Yukawa coupling is below
$1$, so the enhanced $\epsilon_2$ is achieved with
perturbative couplings.
A caveat is, however, that the enhancement nevertheless involves substantial
cancellations among the individual seesaw contributions,
associated with large elements of $R$.
To quantify these cancellations, we define
\begin{align}
    \Delta_{\mathrm{tree}}
    &\equiv
    \frac{\sum_i\|m_\nu^{(i)}\|_F}
         {\|\sum_i m_\nu^{(i)}\|_F}\ ,
    \label{eq:tree_cancellation}\\
    (m_\nu^{(i)})_{\alpha\beta}
    &\equiv
    -\frac{v^2}{M_{N_i}}Y_{i\alpha}Y_{i\beta}\ ,
    \nonumber
\end{align}
where $m_\nu^{(i)}$ is the contribution of $N_i$ to the
tree-level seesaw mass matrix,
$m_\nu^{\mathrm{tree}}=\sum_i m_\nu^{(i)}$
\cite{Minkowski:1977sc,Yanagida:1979as,*Yanagida:1979gs,Gell-Mann:1979vob,Glashow:1979nm,Mohapatra:1979ia},
and $\|\cdot\|_F$ denotes the Frobenius norm.
Values $\Delta_{\mathrm{tree}}\gg1$ indicate that the
norm of the resulting light-neutrino mass matrix is
much smaller than the sum of the norms of its
individual contributions.
For the two benchmarks,
$\Delta_{\mathrm{tree}}\simeq2.3\times10^3$ and
$2.0\times10^4$, respectively,
which indicate fine-tuning of at least
$10^{-3}$ and $10^{-4}$
in the seesaw mechanism, respectively.

It should be also noted that perturbative Yukawa couplings do not guarantee small radiative corrections to the resulting active neutrino mass matrix when
$\Delta_{\mathrm{tree}}\gg1$
\cite{AristizabalSierra:2011mn}.
Reproducing the observed neutrino oscillation parameters precisely
while maintaining a large $\mathcal{E}$ would therefore
require refitting the Yukawa couplings using the seesaw relation with radiative corrections.
However, since our primary aim here is to demonstrate that the CP asymmetry can significantly exceed the Davidson–Ibarra bound, we do not perform a refitting that incorporates radiative corrections in the present analysis.

\begin{table}[tbp]
    \centering
    \caption{
    Benchmark points illustrating enhanced $N_2$ decay asymmetries.
    Both benchmarks use
    $(M_{N_1},M_{N_2},M_{N_3})
    =(5.0\times10^8,5.0\times10^9,5.0\times10^{10})\,\mathrm{GeV}$,
    with $\epsilon_{\mathrm{DI}}\simeq4.9\times10^{-8}$.
    We assume $\mathrm{Br}_{N_2}\simeq1$ and
    $m_\Phi\simeq 2M_{N_2}$.
    The enhancement is evaluated as
    $\mathcal{E}=(\epsilon_2/\epsilon_{\mathrm{DI}})
    (2M_{N_1}/m_\Phi)$.
    The complex angles are defined in Eq.\,\eqref{eq:R_benchmarks}
    and are given in radians.
    The results are evaluated using the displayed benchmark angles.
    Three significant digits are retained in $\operatorname{Im}\phi$
    to reproduce the target enhancements within one percent, and in
    $m_\Phi$ to remain above the $N_2$ pair-production threshold.
    }
    \label{tab:eps2_benchmarks}
    \renewcommand{\arraystretch}{1.15}
    \begin{tabular}{lcc}
        \toprule
        & $\mathrm{BP}_{10^3}$ & $\mathrm{BP}_{10^4}$ \\
        \midrule
        $\mathcal{E}$
        & $9.9\times10^2$ & $1.0\times10^4$ \\
        $\theta$
        & $0$ & $0$ \\
        $\phi$
        & $\pi/4+1.27\,i$ & $\pi/4+2.42\,i$ \\
        $\psi$
        & $3.5\,i$ & $3.5\,i$ \\
        $\epsilon_1$
        & $4.7\times10^{-8}$ & $4.8\times10^{-8}$ \\
        $\epsilon_2$
        & $4.9\times10^{-4}$ & $5.0\times10^{-3}$ \\
        $\epsilon_3$
        & $-2.7\times10^{-5}$ & $-2.7\times10^{-5}$ \\
        $\max_{i,\alpha}|Y_{i\alpha}|$
        & $0.21$ & $0.65$ \\
        $\Delta_{\mathrm{tree}}$
        & $2.3\times10^3$ & $2.0\times10^4$ \\
        $\Gamma_{N_2}\,[\mathrm{GeV}]$
        & $5.3\times10^5$ & $5.3\times10^5$ \\
        \bottomrule
    \end{tabular}
\end{table}

Besides the $N_2$-dominated scenario considered above, a large CP
asymmetry can also arise from resonant enhancement when two
right-handed neutrinos are nearly degenerate
\cite{Pilaftsis:1997jf,Pilaftsis:2003gt}.
When their mass splitting is comparable to their decay widths $\Gamma_N$,
the self-energy contribution can yield $\epsilon\sim\order{1}$.
This requires a very small relative mass splitting,
$\Delta M/M_N\sim\Gamma_N/M_N\ll1$, implying a severe tuning of
the spectrum unless the near degeneracy is protected by a symmetry.
Assuming prompt heavy-neutrino decays, an effective CP asymmetry of order unity
can account for the observed baryon abundance for
$T_R/m_\Phi>\order{10^{-10}}$,
provided that the asymmetry is generated while sphaleron processes
are active, i.e., $T_R >\order{100}$\,GeV.
Non-thermal leptogenesis may therefore remain viable even for
$M_{N_{1,2}}\ll m_\Phi$,
as long as $M_{N_{1,2}}\gg T_R$.
This freedom to separate the right-handed neutrino and the inflaton
mass scales may help alleviate the dark-matter overproduction
from non-thermal processes discussed in the next section.

\section{Non-Thermal Production of Doublet Dark Matter}
\label{sec:nonthermal}
So far, we have considered dark matter production from the thermal bath during inflaton domination, assuming that the inflaton decays into right-handed neutrinos but not directly into dark matter. A suitable assignment of the inflaton's $B-L$ charge can suppress direct dark matter production at leading order.

Even under this assumption, final-state radiation in right-handed-neutrino decays provides an irreducible source of dark matter whenever pair production is kinematically allowed. In addition, energetic particles produced through the inflaton decay chain can generate dark matter  until reheating is completed. In this section, we evaluate these two contributions and examine their implications for the compatibility of heavy pseudo-Dirac doublet dark matter with non-thermal leptogenesis.

\subsection{Dark Matter Emission in Right-Handed Neutrino Decays}

We first consider dark matter production through final-state radiation
in right-handed-neutrino decays. As detailed in
Appendix~\ref{app: dark matter emission}, the lepton or Higgs produced
in the two-body decay can emit an off-shell electroweak gauge boson,
which subsequently splits into a dark matter pair:
\begin{align}
    \bar{N}_R\to\ell+H+V^*\ ,\qquad
    V^*\to\chi+\chi^\dagger\ ,
    \qquad V=B,\ W^a\ .
\end{align}
This process is kinematically allowed for $M_N>2m_D$ and requires
no direct coupling between the inflaton and dark matter.
At leading order, the pair-emission branching fraction is
\begin{equation}
    \mathrm{Br}_D
    \equiv
    \mathrm{Br}(\bar{N}_R\to\ell+H+\chi+\chi^\dagger)
    =
    \frac{
        \Gamma_4(\bar{N}_R\to\ell+H+\chi+\chi^\dagger)
    }{
        \Gamma_N(\bar{N}_R\to\ell+H)
    }\ ,
\end{equation}
where both widths are summed over the relevant doublet components
and charge-conjugate channels.
See the
Appendix~\ref{app: dark matter emission}.
The neutrino Yukawa coupling cancels in this ratio.

\begin{figure}[t]
    \centering
    \includegraphics[width=0.5\linewidth]{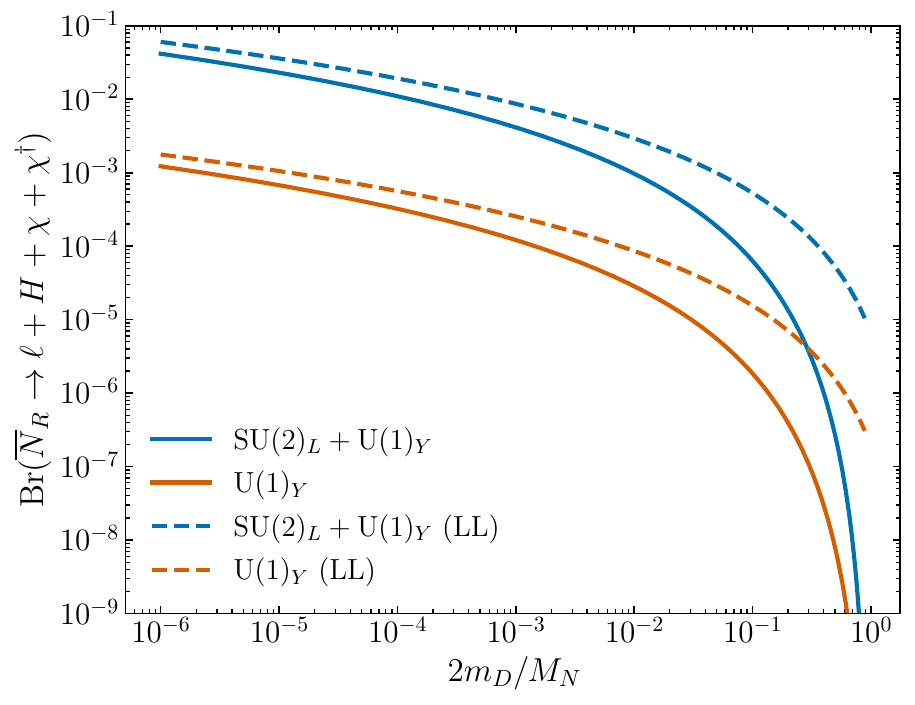}
    \caption{
    Branching fraction for the four-body decay
    $\bar{N}_R\to\ell+H+\chi+\chi^\dagger$
    as a function of $2m_D/M_N$.
    The blue curves include both $\mathrm{SU}(2)_L$ and
    $\mathrm{U}(1)_Y$ gauge interactions, while the red curves
    include only the $\mathrm{U}(1)_Y$ contribution.
    Solid curves show the full numerical results, and dashed
    curves show the corresponding leading-logarithmic (LL)
    approximations valid for $m_D \ll M_N$.
    }
    \label{fig: Branching Ratio}
\end{figure}

Figure~\ref{fig: Branching Ratio} shows the branching fraction
as a function of $2m_D/M_N$.
The solid curves are obtained by numerical phase-space integration
of the squared amplitude which is computed using
FeynArts~\cite{Hahn:2000kx} and
FeynCalc~\cite{Mertig:1990an,Shtabovenko:2016sxi,
Shtabovenko:2020gxv,Shtabovenko:2023idz}. 
For comparison, the dashed curves show the leading-logarithmic
approximation. Including both electroweak gauge interactions,
this approximation takes the simple analytic form in Eq.\,\eqref{eq:leading log},
\begin{equation}    \mathrm{Br}_D\big|_{\mathrm{LL}}
    \sim
    \frac{g^{\prime 4}+3g^4}{12\times192\pi^4}
    \qty(\log\frac{M_N^2}{m_D^2})^3\ ,
    \qquad m_D\ll M_N\ .
\end{equation}
For $g\simeq0.65$ and $g'\simeq0.36$, the full electroweak
branching fraction exceeds $\mathcal{O}(10^{-4})$ for
$2m_D/M_N\lesssim 0.1$.
The leading-logarithmic approximation provides a reasonable
order-of-magnitude estimate for
$2m_D/M_N\lesssim\mathcal{O}(10^{-3})$.%
\footnote{
As discussed in the Appendix~\ref{app: dark matter emission}, the fixed-order perturbative calculation remains reliable provided that the four-body branching fraction is small, i.e., $\Gamma_4/\Gamma_N\ll 1$.
}

This irreducible branching fraction implies that dark matter
production through the inflaton decay chain continues even after
thermal freeze-out during inflaton domination.
Assuming prompt right-handed-neutrino decays and neglecting
subsequent dark matter annihilation, the non-thermal yield
in the instantaneous-reheating approximation is
\begin{align}
    Y_D^{(\mathrm{dec})}
    \sim
    4\,\mathrm{Br}_D
    \frac{n_\Phi}{s_R}\bigg|_{T_R}
    \simeq
    3\qty(\frac{1-r_R}{r_R})\,\mathrm{Br}_D
    \frac{T_R}{m_\Phi}\ .
    \label{eq:YDN}
\end{align}
 The factor of four accounts for the two
right-handed neutrinos produced per inflaton decay and the
two dark-sector particles in each emitted pair.
We have used Eq.\,\eqref{eq: rr} to relate the inflaton and
radiation energy densities at $T=T_R$.

If the injected abundance is sufficiently large, dark matter
undergoes re-annihilation. When this process is efficient,
the final yield is approximately~\cite{Fujii:2001xp}
\begin{align}
    Y_D^{(\mathrm{reann},T_R)}
    \sim
    \frac{H}{s_R\langle\sigma_{\mathrm{eff}}v\rangle}\bigg|_{T_R}
    \sim
    Y_D^{\mathrm{RD,F}}
    \times
    \frac{T_F^{\mathrm{RD}}}{T_R}\ ,
    \label{eq:YD reann TR}
\end{align}
where $Y_D^{\mathrm{RD,F}}$ is the conventional freeze-out
yield in a radiation-dominated universe, and
$T_F^{\mathrm{RD}}\simeq m_D/x^{\mathrm{RD}}_F$, with $x_F^{\mathrm{RD}}\simeq26$,
is the corresponding freeze-out
temperature~\cite{Cirelli:2005uq}.
Using the non-relativistic annihilation cross section in
Eq.\,\eqref{eq:annihilation in NR}, we obtain
\begin{align}
    Y_D^{(\mathrm{reann},T_R)}
    \sim
    10^{-10}
    \qty(\frac{m_D}{10^2\,\mathrm{TeV}})^2
    \qty(\frac{1\,\mathrm{TeV}}{T_R})\ .
\end{align}
The final yield from non-thermal production around
$T\simeq T_R$ is therefore approximately
\begin{align}
    Y_D^{(\mathrm{NT,dec})}
    \sim
    \min\left[
        Y_D^{(\mathrm{dec})},\,
    Y_D^{(\mathrm{reann},T_R)}
    \right]\ .
\end{align}

\subsection{Dark Matter Production in the Thermalization Cascade}

As discussed in the preceding subsection, the dark matter abundance
produced by right-handed-neutrino decays decreases with increasing
inflaton mass at fixed reheating temperature and pair-emission
branching fraction. However, energetic particles produced through
the inflaton decay chain provide an additional source of dark matter
as they thermalize~\cite{Harigaya:2014waa}.

In our setup, each inflaton decays into two right-handed neutrinos,
which subsequently decay into leptons and Higgs bosons.
Assuming prompt right-handed-neutrino decays, the resulting
SM particles have characteristic energies
\begin{align}
    E_{\mathrm{inj}}\sim\frac{m_\Phi}{4}\ .
\end{align}
These particles lose energy through successive gauge-induced
splittings in the thermal bath, as discussed in the previous section.
During this process, collisions between energetic particles and
thermal particles can produce dark matter pairs even when the bath
temperature is much lower than the dark matter mass.

Following Ref.~\cite{Harigaya:2014waa}, we estimate the resulting
dark matter abundance. For a hard particle with energy $E\gg T$,
the center-of-mass energy squared in a collision with a thermal
particle is of order
\begin{align}
    s\sim ET\ .
\end{align}
Dark matter pair production is therefore kinematically allowed for
\begin{align}
    E\gtrsim E_{\mathrm{th}}(T)
    \equiv\frac{m_D^2}{T}\ .
    \label{eq:DM cascade threshold}
\end{align}

At energies of order the production threshold, the pair-production
cross section is parametrically
\begin{align}
    \sigma_{\mathrm{DM}}
    \sim\frac{\alpha_2^2}{m_D^2}\ ,
\end{align}
where we take the $\mathrm{SU}(2)_L$ gauge interaction to dominate
dark matter production. The corresponding production rate per hard
particle is
\begin{align}
    \Gamma_{\mathrm{DM}}(E_{\mathrm{th}})
    \sim
    n_{\mathrm{SM}}\sigma_{\mathrm{DM}}
    \sim
    \frac{\alpha_2^2T^3}{m_D^2}\ ,
    \label{eq:DM production rate cascade}
\end{align}
where $n_{\mathrm{SM}}\sim T^3$ is the number density of thermal
particles.

We assume that the splitting cascade is sufficiently developed,
with a splitting time much shorter than the Hubble time.
The energy injected by inflaton decays is then rapidly transferred
to lower-energy particles. By energy conservation, the cumulative
number of hard particles passing through an energy scale $E$
during a time interval $t\sim H^{-1}$, expressed per unit volume,
is roughly given by,
\begin{align}
    n_h(E)
    \sim
    \frac{m_\Phi}{E}
    n_\Phi\Gamma_\Phi t\ .
    \label{eq:hard particle density cascade}
\end{align}

Each hard particle can produce a dark matter pair through scattering
with the thermal bath before undergoing a further splitting.
The probability for this process is approximately
$\Gamma_{\mathrm{DM}}(E)/\Gamma_{\mathrm{split}}(E)$.
The resulting dark matter number density is therefore estimated as
\begin{align}
    n_D^{\mathrm{(cas)}}(E)
    &\sim
    \frac{\Gamma_{\mathrm{DM}}(E)}
         {\Gamma_{\mathrm{split}}(E)}
    n_h(E)
    \nonumber\\
    &\sim
    \frac{\alpha_2^2T^3}{m_D^2}
    \frac{\sqrt{E}}{\alpha_2^2T\sqrt{T}}
    \frac{m_\Phi}{E}
    n_\Phi\Gamma_\Phi t
    \nonumber\\
    &\sim
    \frac{T^{3/2}}{m_D^2\sqrt{E}}
    m_\Phi n_\Phi\Gamma_\Phi t\ ,
    \label{eq:DM density cascade}
\end{align}
where we have used the splitting rate in
Eq.\,\eqref{eq:splitting_rate}.
Within this estimate, the contribution increases as the hard-particle
energy decreases toward the production threshold.
The dominant contribution therefore comes from
$E\sim E_{\mathrm{th}}$.
Substituting Eq.\,\eqref{eq:DM cascade threshold}, we obtain
\begin{align}
    n_D^{\mathrm{(cas)}}(T)
    \sim
    \frac{m_\Phi T^2}{m_D^3}
    n_\Phi\Gamma_\Phi t\ .
    \label{eq:DM density cascade threshold}
\end{align}

To estimate the final abundance, we use a rough relation
\begin{align}
    \Gamma_\Phi t \sim 
    \frac{\Gamma_\Phi}{H}
    \sim\qty(\frac{T_R}{T})^4\ ,
\end{align}
which follows from the temperature evolution during inflaton
domination derived in the previous section (see Eq.\,\eqref{eq: H in inflaton domination}).

Using $n_\Phi/s_R\sim T_R/m_\Phi$ at $T\simeq T_R$, the contribution
from production at temperature $T>T_R$ to the final dark matter
abundance is estimated as
\begin{align}
m_DY_D^{(\mathrm{cas},\,T)}|_{T_R}
    &\sim
    m_D
    \frac{n_D^{\mathrm{(cas)}}}{n_\Phi}\bigg|_T
    \times \frac{n_\Phi}{s_R}\bigg|_{T_R}
    \nonumber\\
    &\sim
    \frac{T_R^5}{m_D^2T^2}\ .
    \label{eq:DM abundance cascade temperature}
\end{align}
This contribution increases as the production temperature decreases.
The final abundance is therefore dominated by the latest stage
at which pair production remains kinematically allowed.

The kinematic condition for pair production is
\begin{align}
    E_{\mathrm{inj}}T\gtrsim m_D^2\ .
\end{align}
Thus, using $E_{\mathrm{inj}}\sim m_\Phi/4$, we define the
temperature below which the dark matter production is kinematically forbidden,
\begin{align}
    T_{\mathrm{end}}
    \equiv
    \max\qty[
        T_R,\,
        \frac{4m_D^2}{m_\Phi}
    ]\ .
    \label{eq:DM cascade dominant temperature}
\end{align}
This estimate assumes $T_{\mathrm{end}}<T_{\mathrm{max}}$
and a sufficiently developed cascade.

Substituting $T_{\mathrm{end}}$ into
Eq.\,\eqref{eq:DM abundance cascade temperature}, we obtain
\begin{align}
    m_DY_D^{\mathrm{(cas)}}
    &\sim
    \frac{T_R^5}{m_D^2T_{\mathrm{end}}^2}
    \nonumber\\
    &\sim
    \begin{cases}
        \displaystyle
        \frac{T_R^3}{m_D^2}\ ,
        &
        \displaystyle
        T_R\geq\frac{4m_D^2}{m_\Phi}\ ,
        \\[3mm]
        \displaystyle
        \frac{T_R^3}{m_D^2}
        \qty(\frac{m_\Phi T_R}{4m_D^2})^2\ ,
        &
        \displaystyle
        T_R<\frac{4m_D^2}{m_\Phi}\ .
    \end{cases}
\end{align}
In the first case, production continues until reheating
is completed, and the abundance is independent of
the inflaton mass.
In the second case, production ends earlier, and the
resulting abundance is diluted by subsequent entropy
production.
At fixed $T_R$ and $m_D$, a lighter inflaton causes
production to end at a higher temperature, leading
to a smaller final abundance.

If re-annihilation is efficient when production ends, the final
yield, including dilution between $T_{\mathrm{end}}$ and $T_R$,
is estimated as
\begin{align}
Y_D^{(\mathrm{reann},T_\mathrm{end})}|_{T_R}
    &\sim
\frac{H}{\rho_\Phi \expval{\sigma_{\mathrm{eff}} v }}\bigg|_{T_\mathrm{end}}\times
\frac{\rho_\Phi}{s_R}\bigg|_{T_R}
\sim Y_{D}^{(\mathrm{reann},T_R)}\times\qty(\frac{T_R}{T_\mathrm{end}})^4 \ ,
    \label{eq: YD reann Tend}
\end{align}
where $Y_{D}^{(\mathrm{reann},T_R)}$
is given in Eq.\,\eqref{eq:YD reann TR}.
The final abundance is therefore approximately given by,
\begin{align}
    Y_D^{\mathrm{(NT,cas)}}
    \sim
    \min\left[
Y_D^{(\mathrm{cas},T_\mathrm{end})},\,        Y_D^{(\mathrm{reann},T_\mathrm{end})}
    \right]\ .
\end{align}

\subsection{Summary of Non-Thermal Dark Matter Production}

We now summarize the two sources of non-thermal dark matter
production discussed above.
When the
re-annihilation is inefficient, four-body decays of right-handed
neutrinos give
\begin{align}
    m_DY_D^{(N)}
    \sim
    3\qty(\frac{1-r_R}{r_R})\,\mathrm{Br}_D\,
    \frac{m_DT_R}{m_\Phi}\ ,
    \label{eq:DM abundance summary decay}
\end{align}
where $\mathrm{Br}_D$ is the branching fraction for dark matter
pair emission shown in Fig.\,\ref{fig: Branching Ratio}.
The contribution from the thermalization cascade is, on the other hand, given by
\begin{align}
    m_DY_D^{\mathrm{(cas)}}
    \sim
    \begin{cases}
        \displaystyle
        0.1\,\mathrm{GeV}
        \qty(\frac{T_R}{1\,\mathrm{TeV}})^3
        \qty(\frac{10^2\,\mathrm{TeV}}{m_D})^2\ ,
        &
        \displaystyle
        m_\Phi\gtrsim\frac{4m_D^2}{T_R}\ ,
        \\[5mm]
        \displaystyle
        6\times10^{-7}\,\mathrm{GeV}
        \qty(\frac{T_R}{1\,\mathrm{TeV}})^5
        \qty(\frac{10^2\,\mathrm{TeV}}{m_D})^4
        \qty(\frac{m_\Phi}{m_D})^2\ ,
        &
        \displaystyle
        m_\Phi\lesssim\frac{4m_D^2}{T_R}\ .
    \end{cases}
    \label{eq: cascade}
\end{align}

If re-annihilation is efficient, it reduces the abundance
relative to these production estimates, as described in
Eqs.\,\eqref{eq:YD reann TR} and \eqref{eq: YD reann Tend}.
Notice that Eq.\,\eqref{eq: cascade} provides only a rough estimate, and the results below should be interpreted at the
order-of-magnitude level.

\begin{figure*}[t]
    \centering
    \begin{subfigure}[t]{0.45\textwidth}
        \centering
        \includegraphics[width=\linewidth]
        {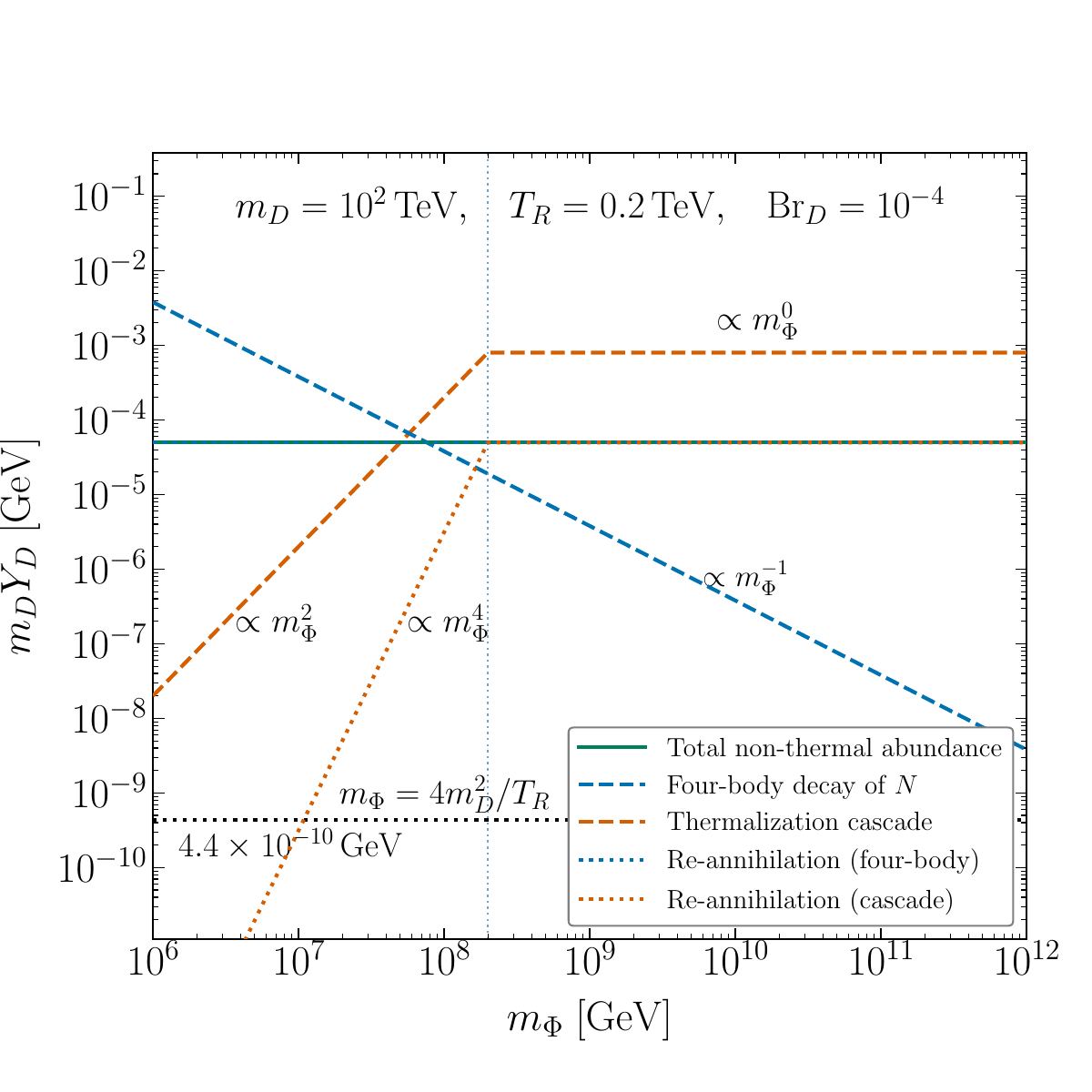}
        \caption{
        $m_D=10^2\,\mathrm{TeV}$,
        $T_R=0.2\,\mathrm{TeV}$.
        }
        \label{fig:nonthermal_100TeV}
    \end{subfigure}
    \hfill
    \begin{subfigure}[t]{0.45\textwidth}
        \centering
        \includegraphics[width=\linewidth]
        {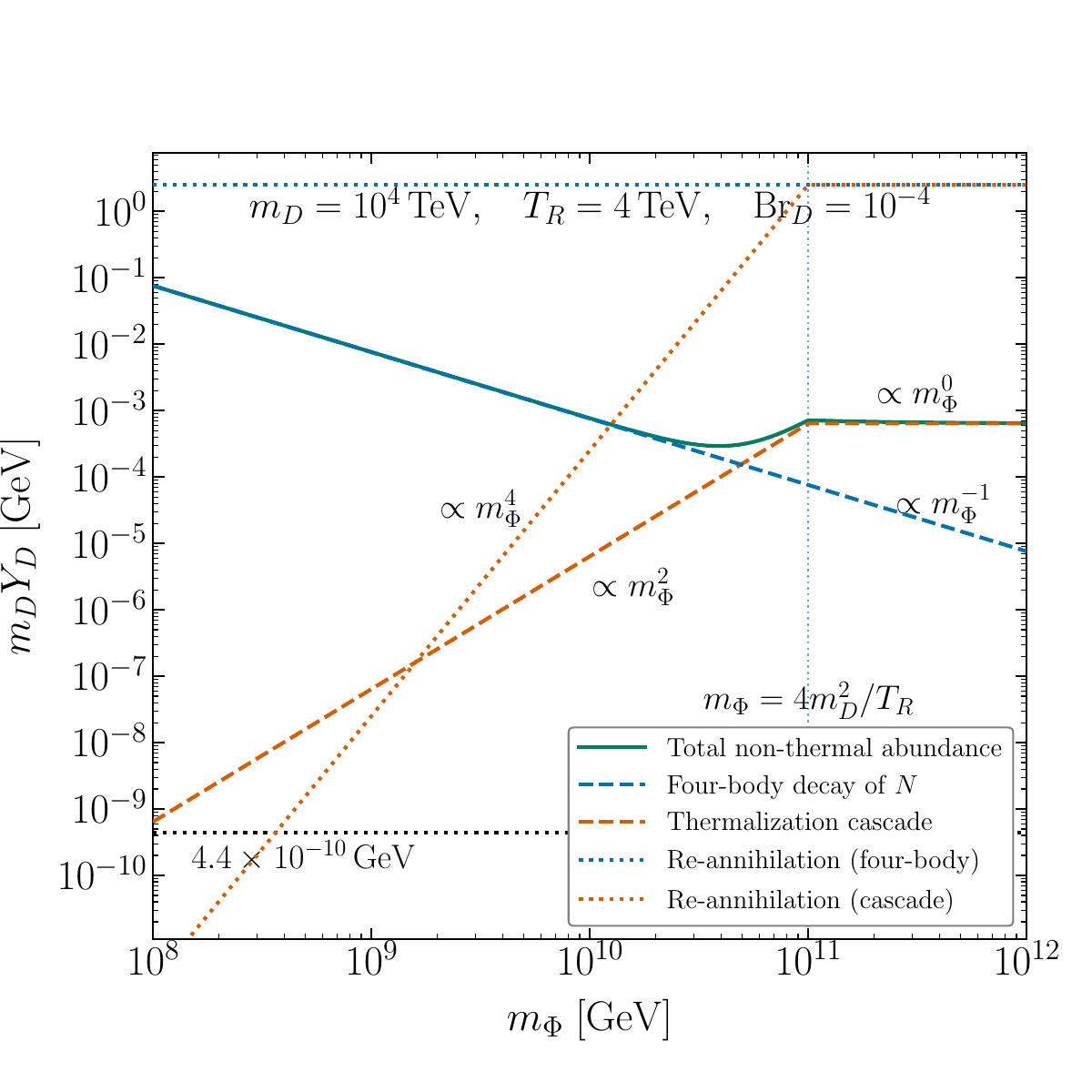}
        \caption{
        $m_D=10^4\,\mathrm{TeV}$,
        $T_R=4\,\mathrm{TeV}$.
        }
        \label{fig:nonthermal_10000TeV}
    \end{subfigure}

    \caption{
    Non-thermal dark matter abundance as a function of the
    inflaton mass $m_\Phi$ for two benchmark points,
    with $\mathrm{Br}_D=10^{-4}$.
    The blue and red dashed curves show the decay and
    cascade contributions before re-annihilation, respectively.
    The blue and red dotted curves show the re-annihilation
    estimates for production ending at $T_R$ and
    $T_{\mathrm{end}}$, respectively, including subsequent
    entropy dilution in the latter case.
    The green solid curve shows the estimated final abundance
    including re-annihilation.
    The horizontal black-dotted line indicates the observed
    dark matter abundance, while the vertical dotted line
    marks $m_\Phi=4m_D^2/T_R$.
    }
    \label{fig:nonthermal_DM}
\end{figure*}

Figure\,\ref{fig:nonthermal_DM} illustrates these contributions
for $(m_D,T_R)=(10^2\,\mathrm{TeV},0.2\,\mathrm{TeV})$ and
$(10^4\,\mathrm{TeV},4\,\mathrm{TeV})$, where the reheating temperature is taken to be $T_R = T_R^{(\mathrm{th})}(m_D)$.
We fix $\mathrm{Br}_D=10^{-4}$ to illustrate their different
dependences on $m_\Phi$.
The four-body decay contribution decreases with increasing $m_\Phi$,
whereas the cascade contribution increases below
$m_\Phi\sim4m_D^2/T_R$ and becomes independent
of $m_\Phi$ above this threshold.

For the first benchmark, re-annihilation plays an important
role in determining the final abundance.
As a result, the final abundance corresponds to the value after re-annihilation, shown by the dotted lines.
For the second benchmark, it has little effect on the total
abundance, which is approximately given by the sum of four-body decay and cascade production contributions.
In both cases, the estimated final abundance remains well
above the observed dark matter abundance.

Avoiding overproduction therefore requires suppressing both
sources.
A lower inflaton mass suppresses cascade production in the
regime $m_\Phi\lesssim4m_D^2/T_R$.
At fixed $\mathrm{Br}_D$, however, it increases the decay
contribution, so the four-body branching fraction must
also be reduced.

Let us consider the $N_2$-dominated scenario discussed in
Subsec.\,\ref{sec:Numerical Analysis}.
In this case, dark matter pair emission
in $N_2$ decays can be kinematically forbidden by taking
$M_{N_2}<2m_D$.
The benchmarks considered in the previous section, however, assumed
$m_\Phi\simeq2M_{N_2}$, which approximately maximizes the
non-thermal leptogenesis yield for a given CP asymmetry.
Thus, taking $m_\Phi\gg2M_{N_2}$ requires a larger
$\epsilon_2/\epsilon_{\mathrm{DI}}$ to maintain the same
enhancement factor $\mathcal{E}$ at fixed
right-handed-neutrino mass ratios.
This requires
stronger cancellations among the seesaw contributions and
hence a greater degree of fine-tuning.

Alternatively, $M_N<2m_D$ can be accommodated in resonant
leptogenesis, where a sufficiently large CP asymmetry can
arise from the near degeneracy of two right-handed neutrinos.
In this case, $M_N\ll m_\Phi$ does not by itself suppress
the decay CP asymmetry. 
In this case, 
the leptogenesis provides 
sufficient baryon asymmetry 
for  $T_R/m_\Phi \gg \order{10^{-10}}$.

\begin{figure*}[t]
    \centering
    \begin{subfigure}[t]{0.45\textwidth}
        \centering
        \includegraphics[width=\linewidth]
        {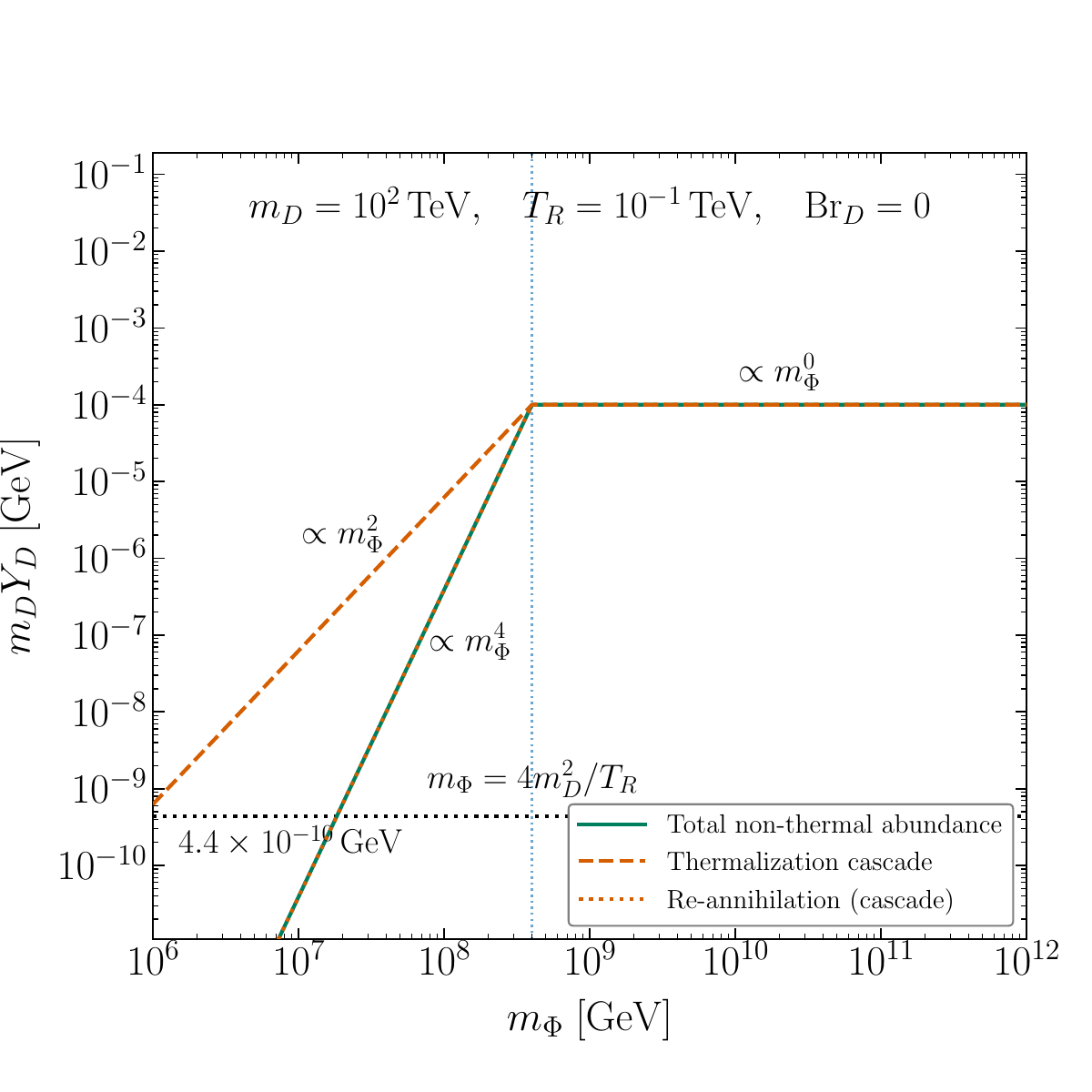}
        \caption{
        $m_D=10^2\,\mathrm{TeV}$,
        $T_R=0.1\,\mathrm{TeV}$.
        }
        \label{fig:nonthermal_100TeV_BrD0}
    \end{subfigure}
    \hfill
    \begin{subfigure}[t]{0.45\textwidth}
        \centering
        \includegraphics[width=\linewidth]
        {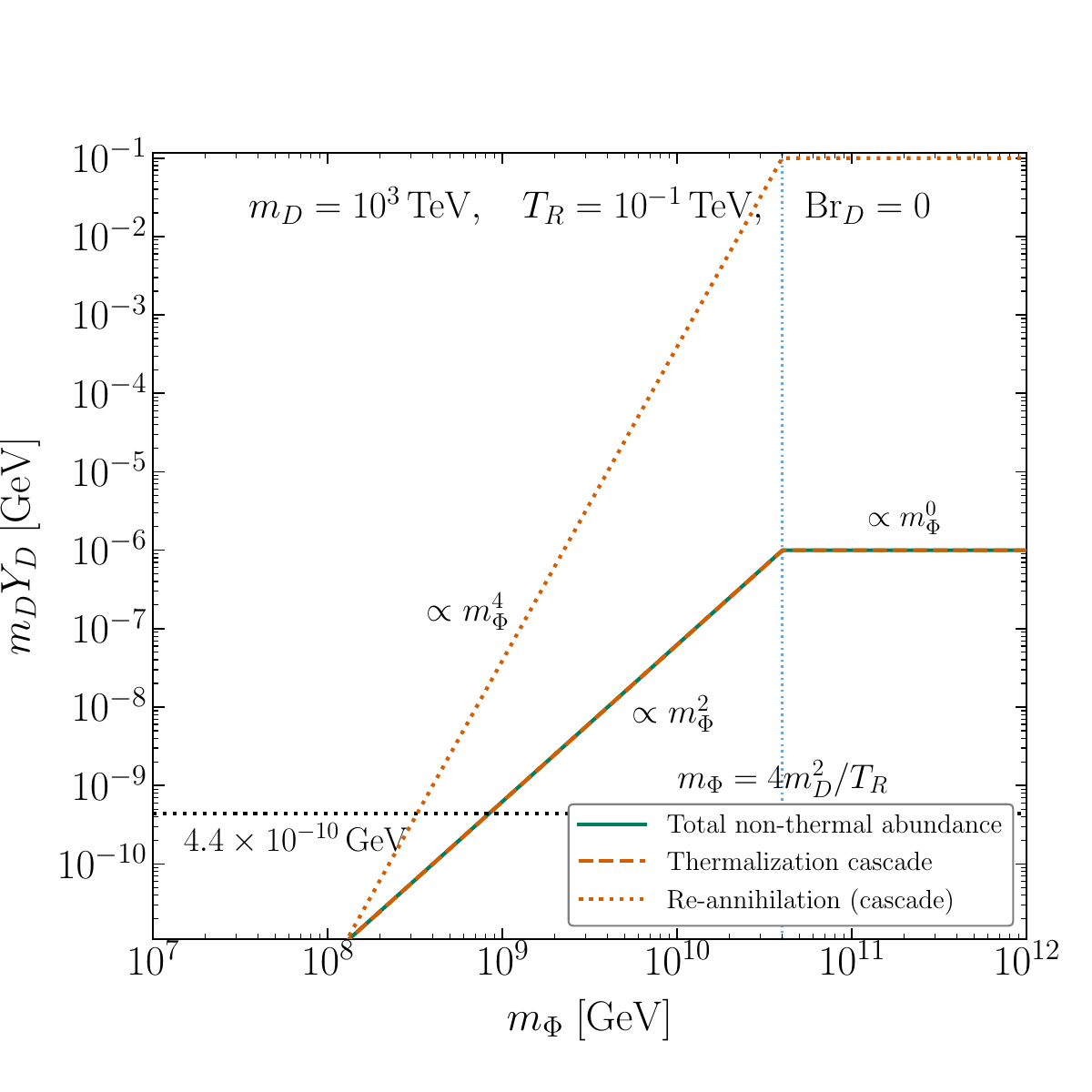}
        \caption{
        $m_D=10^3\,\mathrm{TeV}$,
        $T_R=0.1\,\mathrm{TeV}$.
        }
        \label{fig:nonthermal_1000TeV_BrD0}
    \end{subfigure}

    \caption{
    Non-thermal dark matter abundance as a function of the
    inflaton mass $m_\Phi$ for
    $m_D=10^2\,\mathrm{TeV}$ and $10^3\,\mathrm{TeV}$,
    with $T_R=0.1\,\mathrm{TeV}$ and $\mathrm{Br}_D=0$.
    The curve conventions are the same as in
    Fig.\,\ref{fig:nonthermal_DM}.
    }
    \label{fig:nonthermal_DM_BrD_0}
\end{figure*}

With the above difficulties in mind,
let us assume the scenarios with $M_N<2m_D$,
and show the dark matter abundance for $\mathrm{Br}_D=0$ in Fig.\,\ref{fig:nonthermal_DM_BrD_0}.
Here, we take two benchmark points,
$(m_D,T_R)=(10^2\,\mathrm{TeV},0.1\,\mathrm{TeV})$ and
$(10^3\,\mathrm{TeV},0.1\,\mathrm{TeV})$. 
We choose $T_R=0.1\,\mathrm{TeV}$ to suppress the thermalization cascade contribution (see Eq.\,\eqref{eq: cascade}), where
thermally produced dark matter abundance is lower than the observed dark matter density.

Matching the final cascade-produced abundance, including
re-annihilation, to the observed dark matter abundance gives
\begin{align}
    m_\Phi \sim \max\Bigg[
    &10^6\,\mathrm{GeV}
    \qty(\frac{0.1\,\mathrm{TeV}}{T_R})^{5/2}
    \qty(\frac{m_D}{10^2\,\mathrm{TeV}})^3\ ,
    \nonumber\\
    &2\times10^7\,\mathrm{GeV}
    \qty(\frac{0.1\,\mathrm{TeV}}{T_R})^{3/4}
    \qty(\frac{m_D}{10^2\,\mathrm{TeV}})^{5/4}
    \Bigg]\ .
    \label{eq: mPhi for DM abundance}
\end{align}
The first entry applies when re-annihilation is inefficient,
and the second when it is efficient.
This expression assumes $T_\mathrm{end}>T_R$.
The inferred values of $m_\Phi$ and $\epsilon$ should be regarded as order-of-magnitude estimates. In particular, the re-annihilation result is sensitive to the treatment of the end of cascade production, since the high-energy tail of the thermal bath allows pair production to persist below the nominal $T_{\rm end}$.
Assuming efficient sphaleron conversion and negligible washout,
we obtain the required effective CP asymmetry from
Eq.\,\eqref{eq: YB} as
\begin{align}
    \epsilon \sim \max\Bigg[
    &2\times10^{-6}
    \qty(\frac{0.1\,\mathrm{TeV}}{T_R})^{7/2}
    \qty(\frac{m_D}{10^2\,\mathrm{TeV}})^3\ ,
    \nonumber\\
    &4\times10^{-5}
    \qty(\frac{0.1\,\mathrm{TeV}}{T_R})^{7/4}
    \qty(\frac{m_D}{10^2\,\mathrm{TeV}})^{5/4}
    \Bigg]\ .
\end{align}

These estimates suggest that non-thermal leptogenesis may be
compatible with heavy pseudo-Dirac doublet dark matter
provided that the following conditions are satisfied.
\begin{itemize}
    \item Dark matter pair emission in the decays of the relevant
    right-handed neutrinos is kinematically forbidden.
    \item Cascade production is sufficiently suppressed by a low
    reheating temperature and a suitable inflaton mass.
\end{itemize}
A viable realization also requires a sufficiently large CP
asymmetry, efficient sphaleron conversion, and negligible washout.
Achieving the required CP asymmetry in the scenarios discussed
above, however, typically requires fine-tuning among the seesaw
contributions or in the right-handed neutrino mass spectrum.

\section{Conclusions}
\label{sec:conclusions}
In this paper, we have investigated a cosmological scenario accommodating heavy pseudo-Dirac electroweak doublet dark matter and non-thermal leptogenesis, motivated by the recent LZ results. By solving the coupled Boltzmann equations during reheating, we have shown that entropy production following dark matter freeze-out can sufficiently dilute the excessive thermal relic abundance. For dark matter masses of $10^2$–$10^4\,\mathrm{TeV}$, the observed abundance can be reproduced with reheating temperatures of order a TeV. 

Such low reheating temperatures, however, pose a challenge to leptogenesis. 
Even in non-thermal leptogenesis with the conventional $N_1$-dominated scenario, the Davidson–Ibarra bound implies an insufficient baryon asymmetry. The required CP asymmetry exceeds this bound by approximately $10^3$–$10^4$, even when the inflaton mass lies near the right-handed-neutrino pair threshold. 
We have presented $N_2$-dominated benchmarks illustrating that sufficient enhancement is possible with perturbative Yukawa couplings, although substantial cancellations among the seesaw contributions are required. 

We have further examined non-thermal dark matter production through right-handed-neutrino four-body decays.
We also examine the non-thermal dark matter production through the thermalization cascade. 
Both processes can generate dark matter even after freeze-out.
Increasing the inflaton mass suppresses the four-body decay contribution, which however enhances cascade production. Thus, varying the inflaton mass alone does not resolve the overproduction problem. 

Our estimates do not exclude a regime in which right-handed-neutrino four-body decays are kinematically forbidden, $M_N<2m_D$. However, reproducing both observed abundances would require a sufficiently low reheating temperature, an appropriate inflaton mass, enhanced CP violation from tuned seesaw cancellations or nearly degenerate right-handed neutrinos. 
Whether these requirements can be satisfied simultaneously calls for a more detailed analysis.

Finally, we comment on the compatibility of the low-reheating-temperature leptogenesis scenario recently discussed in Ref.\,\cite{Garcia:2026ulw} with heavy pseudo-Dirac doublet dark matter. 
In the radiation-like fermionic reheating scenario studied there, the inflaton has an effective equation of state $w_\Phi=1/3$. 
This framework allows successful non-thermal leptogenesis even at reheating temperatures well below the electroweak scale. 
However, entropy dilution is substantially weaker than in the matter-like reheating scenario considered in this work. This makes it difficult to sufficiently suppress the thermal relic abundance of a pseudo-Dirac electroweak doublet with a mass of $10^2$–$10^4\,\mathrm{TeV}$.

\section{Acknowledgments}
This work is supported by Grant-in-Aid for Scientific Research from the Ministry of Education, Culture, Sports, Science, and Technology (MEXT), Japan,  24K23938, 25H00644, 26K07102
(M.I.)
and by World Premier International Research Center Initiative (WPI),
MEXT, Japan.  

\appendix
\section{Annihilation Cross Section and Sommerfeld Enhancement factor}
\label{app: cross section}
The velocity dependence of the annihilation cross section of the pseudo-Dirac doublet into the SM particles is given by,
\begin{align}
    \sigma_\mathrm{eff} v 
    &=\frac{1}{8}
    \qty(\sigma_\mathbf{1} v + 3 \sigma_\mathbf{3} v)\ , 
    \label{eq: sigma eff v}\\
    \sigma_\mathbf{1}v & =   \frac{(3g^4+g'^4)I_0
    (\beta)}{128\pi s}
    +
    \frac{41g'^4}{64\pi s}
    \left(1-\frac{\beta^2}{3}\right)\ , \\  
    \sigma_\mathbf{3}v&=     \frac{
        g^4 I_2(\beta)
        +g^2g'^2 I_0(\beta)
    }{64\pi s}
    +
    \frac{25g^4}{64\pi s}
    \left(1-\frac{\beta^2}{3}\right)\ .
\end{align}
Here, 
$s$ denotes the 
Mandelstam variable, and their relative velocity is given by,
\begin{align}
    \beta
    =\sqrt{1-\frac{4m_D^2}{s}}\ ,
    \qquad
    v=2\beta\ ,
 \end{align}
 in the center of mass frame. 
 The functions $I_0(\beta)$ and 
 $I_2(\beta)$ are given by,
 \begin{align}
    I_0(\beta)
    &=
    2\left[
        -4+2\beta^2
        +\frac{3-\beta^4}{\beta}L_\beta
    \right] \ ,
    \label{eq:I0-rate}
    \\
    I_2(\beta)
    &=
    -20+\frac{32}{3}\beta^2
    +\frac{2(\beta^4-4\beta^2+5)}{\beta}L_\beta
    \label{eq:I2-rate} \ , \\
    L_\beta &= \log\frac{1+\beta}{1-\beta}\ .
\end{align}
respectively.
We have decomposed the initial state according to the $\mathrm{SU}(2)_L$ representation,
\begin{align}
\mathbf{2}\otimes\bar{\mathbf{2}}
= \mathbf{1}\oplus\mathbf{3}\ .
\end{align}
This decomposition will be used later to show that the Sommerfeld enhancement is small.

The effective cross section
reproduces the reduced annihilation cross section $\hat{\sigma}_A$ in Ref.\,\cite{Cirelli:2009uv} through%
\begin{align}
    \sigma_\mathrm{eff} v = \frac{1}{16}\frac{1}{sv}\hat{\sigma}_A\ .
\end{align}
The thermal average of the cross section is given by,
\begin{align}
    \langle\sigma_{\mathbf R}v\rangle(T)
    =
    \frac{1}{16m_D^4T K_2(x)^2}
    \int_{4m_D^2}^{\infty}ds\,
    s^{3/2}\beta\,
    \sigma_{\mathbf{R}} v\,
    K_1\!\left(\frac{\sqrt{s}}{T}\right)
    \ ,
    \label{eq:thermal-irrep-unexpanded}
\end{align}
for $\mathbf{R} = \mathbf{1}, \mathbf{3}$. (See Ref.\,\cite{Gondolo:1990dk}.)
In the non-relativistic limit the effective cross section in Eq.\,\eqref{eq: sigma eff v} reduces to Eq.\,\eqref{eq:annihilation in NR}.

We next estimate the Sommerfeld correction relevant for thermal
freeze-out~\cite{Hisano:2003ec,Hisano:2004ds}. For the heavy doublet considered here, the freeze-out
temperature $T_F=m_D/x_F$, with $x_F=\mathcal{O}(10)$, is well above the electroweak symmetry breaking scale. We therefore work in the
electroweak-symmetric phase.

For a doublet--antidoublet pair with hypercharges $\pm1/2$,
electroweak gauge-boson exchange gives the potentials
\begin{align}
    V_{\mathbf{1}}(r)
    =-\frac{3\alpha_2+\alpha_Y}{4r}\ ,
    \qquad
    V_{\mathbf{3}}(r)
    =\frac{\alpha_2-\alpha_Y}{4r}\ ,
    \label{eq:doublet-coulomb-potentials}
\end{align}
where $\alpha_2=g^2/(4\pi)$ and $\alpha_Y=g'^2/(4\pi)$.
Thus, the singlet channel is attractive, whereas the triplet
channel is repulsive. Writing
$V_{\mathbf{R}}(r)=-\alpha_{\mathbf{R}}/r$, we have
\begin{align}
    \alpha_{\mathbf{1}}
    =\frac{3\alpha_2+\alpha_Y}{4}\ ,
    \qquad
    \alpha_{\mathbf{3}}
    =\frac{\alpha_Y-\alpha_2}{4}\ .
\end{align}
The corresponding Sommerfeld factor for $s$-wave annihilation
is~\cite{Cirelli:2009uv}
\begin{align}
    S_{\mathbf{R}}(v)
    &=
    \frac{2\pi\alpha_{\mathbf{R}}/v}
    {1-\exp[-2\pi\alpha_{\mathbf{R}}/v]}\ ,
    \label{eq:doublet-sommerfeld}
\end{align}
where $v$ denotes the relative velocity.
In particular, $S_{\mathbf{1}}>1$ and $S_{\mathbf{3}}<1$.

To estimate the correction in the non-relativistic regime,
we retain the leading $s$-wave annihilation coefficients.
Using $I_0(0)=4$ and $I_2(0)=0$, the cross sections above give
\begin{align}
    a_{\mathbf{1}}
    &\equiv \lim_{v\to0}\sigma_{\mathbf{1}}v
    =\frac{6g^4+43g'^4}{256\pi m_D^2}\ ,
    \\
    a_{\mathbf{3}}
    &\equiv \lim_{v\to0}\sigma_{\mathbf{3}}v
    =\frac{25g^4+4g^2g'^2}{256\pi m_D^2}\ .
\end{align}
For a Maxwell--Boltzmann velocity distribution, the thermal
average of each Sommerfeld factor is
\begin{align}
    \overline{S}_{\mathbf{R}}(x)
    &=
    \frac{x^{3/2}}{2\sqrt{\pi}}
    \int_0^\infty dv\,
    v^2 e^{-xv^2/4}S_{\mathbf{R}}(v)\ ,
    \qquad x=\frac{m_D}{T}\ .
    \label{eq:thermal-sommerfeld-doublet}
\end{align}
The correction to the effective annihilation rate in this
approximation is therefore
\begin{align}
    \overline{S}_{\mathrm{eff}}(x)
    &\equiv
    \frac{
        a_{\mathbf{1}}\overline{S}_{\mathbf{1}}(x)
        +3a_{\mathbf{3}}\overline{S}_{\mathbf{3}}(x)
    }{
        a_{\mathbf{1}}+3a_{\mathbf{3}}
    }\ .
    \label{eq:effective-sommerfeld-doublet}
\end{align}
For $g=0.65$ and $g'=0.36$,  we find
$\overline{S}_\mathrm{eff}$ is about $1$\,\% effect
for $x=10$--$30$,
where the enhancement in the singlet channel is canceled by the suppression in the triplet channel.

\section{Energy Density Fraction at Reheating Temperature}

\label{app: Energy Density Ratio}
In this appendix, we estimate the radiation energy density fraction $r_R$ at the reheating temperature, as used in Sec.\,\ref{sec: Thermal Abundance}. 

The Boltzmann equations governing the evolution of the
inflaton and radiation energy densities,
Eqs.~\eqref{eq:inflaton_development} and
\eqref{eq:radiation_development}, can be rewritten in
terms of dimensionless variables as
\begin{align}
    \frac{d}{dN} X_\Phi + 3 X_\Phi &= - \frac{X_\Phi}{\sqrt{X_\Phi + X_R}}\ , \\
    \frac{d}{dN} X_R + 4 X_R &=  \frac{X_\Phi}{\sqrt{X_\Phi + X_R}}\ ,
\end{align}
where $X_j \equiv \rho_j/(3M_{\rm Pl}^2\Gamma_\Phi^2)$
$(j=\Phi,R)$, $N \equiv \ln(a/a_i)$,
and $a_i$ denotes the initial scale factor.

To investigate the evolution of the radiation fraction,
we introduce the variables
\begin{align}
    \Xi \equiv \frac{X_R}{X_\Phi+X_R}\ , \qquad 
    \xi \equiv \frac{1}{\sqrt{X_\Phi+X_R}}\ ,
\end{align}
where $0\leq \Xi \leq 1$, and $\xi$ is an increasing function of $N$ by definition.
In terms of these variables, the Boltzmann equations reduce to the following autonomous dynamical system:
\begin{align}
    \frac{d\Xi}{dN}
    = (1-\Xi)(\xi-\Xi)\ , \qquad
    \frac{d\xi}{dN}
    = \frac{1}{2}(3+\Xi)\xi\ ,
\end{align}
with the initial condition $\Xi_i=0$, and $\xi_i=\Gamma_\Phi/H_i$, where $H_i$ is the initial Hubble scale.
Eliminating $N$, we obtain
\begin{align}
    \frac{d\Xi}{d\xi}
    = f(\xi,\Xi)\ ,\qquad
    f(\xi,\Xi)
    =\frac{2(1-\Xi)(\xi-\Xi)}
    {\xi(3+\Xi)}\ .
\label{eq:omega_h_evolution}
\end{align}

From the definition of the reheating temperature
in Eq.\,\eqref{eq:TRH_Def}, we have $X_R=1/9$ at $T=T_R$.
Since $X_R=\Xi/\xi^2$, this condition gives
\begin{align}
    r_R = \left.\Xi\right|_{T_R}=\frac{\xi^2}{9}\bigg|_{T_R}\ .
    \label{eq: rR in terms of xi}
\end{align}
In the following, we estimate $r_R$  using inequalities.

For this purpose, we introduce the following auxiliary function $K(\xi)$,
\begin{align}
    K(\xi) = \frac{2\xi}{5+\xi}\ .
\end{align}
This auxiliary function satisfies,
\begin{align}
     f(\xi,K(\xi)) < \frac{dK}{d\xi}\ ,
    \qquad  \Xi(\xi_i) < K(\xi_i) \ .
\end{align}
Then, according to the comparison theorem for ordinary differential equations, 
we find 
\begin{align}
    \Xi(\xi) < K(\xi)\ ,
    \label{eq: inequality 1}
\end{align} 
for all $\xi_i \leq \xi$.

In addition, to derive a lower bound on $\Xi$, we introduce
a new variable $G(\xi)\equiv-\ln(1-\Xi(\xi))$.
The evolution equation for $G(\xi)$ is given by
\begin{align}
    \frac{dG}{d\xi}
    = \frac{2(\xi-\Xi)}{\xi(3+\Xi)}\ .
\end{align}
Substituting the inequality in Eq.\,\eqref{eq: inequality 1}
into the above equation,
we obtain
\begin{align}
    \frac{2}{5} \leq \frac{dG}{d\xi} \ .
    \label{app: dGdhandOmega}
\end{align}
Integrating this inequality from $\xi_i$ to $\xi$
with $G(\xi_i)=0$, we obtain
\begin{align}
    \frac{2}{5}(\xi-\xi_i) < G(\xi) \ .
\end{align}
Using $\Xi=1-e^{-G}$, we finally obtain
\begin{align}
    1-e^{-\frac{2}{5}(\xi-\xi_i)}
    <\Xi(\xi)<\frac{2\xi}{5+\xi}\ ,
\end{align}
for $\xi_i<\xi$.
By using Eq.\,\eqref{eq: rR in terms of xi}
at $T=T_R$, we can determine the 
range of $\xi$ at $T_R$.
In the limit $\xi_i\to0$ (i.e., $H_i \gg \Gamma_\Phi$), 
we obtain the bounds of $r_R$
\begin{align}
    0.61 \lesssim r_R \lesssim 0.65\ .
\end{align}

\section{Freeze-Out Temperature}
\label{app: freeze-out temperature}
In this appendix, we examine the dependence of $x_F$ on the mass $m_D$ for $T_R = T_R^{(\mathrm{th})}(m_D)$.
The Boltzmann equation given in Eq.\,\eqref{eq:Boltzmann-Eq-x} can be rewritten as 
\begin{align}
    \frac{d}{dx}(x^8 n_D) =-\frac{8\langle\sigma_\mathrm{eff} v\rangle}{3H}x^7
    \left(n_D^2-n_{D,\mathrm{eq}}^2\right)\ .
\end{align}
Before freeze-out, the number density $n_D$ closely tracks its equilibrium value $n_{D,\rm{eq}}$.
To estimate the mass dependence of $x_F$,
expand $n_D$ as $n_D = n_{D,\rm{eq}} + \varDelta$.
At the early time, we may neglect $d(x^8 \varDelta)/dx$, obtaining
\begin{align}
    \frac{d}{dx}\left(x^8 n_{D,\rm{eq}}\right) 
    \simeq -\frac{8\langle\sigma_\mathrm{eff} v\rangle}{3H}x^7
    \varDelta \left(2n_{D,\rm{eq}} + \varDelta\right)\ ,
\end{align}
for $x\simeq x_F$.
Substituting $\varDelta = c n_{D,\rm{eq}}$
and using the non-relativistic equilibrium density 
$n_{D,\rm{eq}} \propto x^{-3/2}e^{-x}$, we find
\begin{align}
    \left(\frac{13}{2} - x\right)
    \simeq -\frac{8\langle\sigma_\mathrm{eff} v\rangle}{3H}
    c(2+c)n_{D,\rm{eq}}\ .
\end{align}

We estimate the onset of freeze-out as the point at which the
departure from equilibrium becomes comparable to the equilibrium
density, corresponding to $c=\mathcal{O}(1)$.
Evaluating the above relation at $x=x_F$ therefore yields
\begin{align}
    n_{D,\mathrm{eq}}(x_F)\langle\sigma_\mathrm{eff} v\rangle_F
    \sim
    \left(x_F-\frac{13}{2}\right)H_F\ ,
    \label{eq:at xF}
\end{align}
up to a numerical factor of order unity, where
$H_F\equiv H(x_F)$ and
$\langle\sigma_\mathrm{eff} v\rangle_F\equiv\langle\sigma_\mathrm{eff} v\rangle(x_F)$.

For $\langle\sigma_\mathrm{eff} v\rangle\propto m_D^{-2}$, the freeze-out condition implies
\begin{align}
    \left(x_F-\frac{13}{2}\right)x_F^{-5/2}e^{x_F}
    \propto T_R^2m_D^{-3}\ ,
\end{align}
where we have used $H_F=H_R(x_R/x_F)^4$,
$H_R\propto T_R^2$, and $x_R=m_D/T_R$.
On the other hand, for a given $x_F$,
the dark matter abundance $m_D Y_D$ is given in Eq.\,\eqref{eq: YD}.
Thus, by requiring that the predicted reproduces the observed dark matter abudnance, we find 
\begin{align}
    T_R^{(\rm{th})} \propto m_{D}^{1/3} x_F^{-4/3}\ .
\end{align}
Combining these relations, we obtain
\begin{align}
    \left(x_F-\frac{13}{2}\right)x_F^{1/6}e^{x_F}
    \propto m_D^{-7/3}\ .
\end{align}
Thus, the logarithmic differentiation along the contour of fixed relic abundance then yields
\begin{align}
    \dv{\ln T_R^{(\rm{th})}}{\ln m_D}
    &=
    \frac{1}{3}
    -\frac{4}{3}
    \frac{d \ln x_F}{d\ln m_D}\ , \\
    \dv{\ln x_F}{\ln m_D}
    &= -\frac{7}{3}
    \qty(
        x_F+\frac{1}{6}+\frac{2x_F}{2x_F-13}
    )^{-1}\ .
\end{align}
For $x_F\simeq10$--$30$, the logarithmic slope
of $T_R$ is approximately $0.57$--$0.43$,
which accounts for the approximate scaling
$T_R^{(\rm{th})}\propto m_D^{1/2}$.

\begin{figure}[tbp]
    \centering    \includegraphics[width=0.55\linewidth]{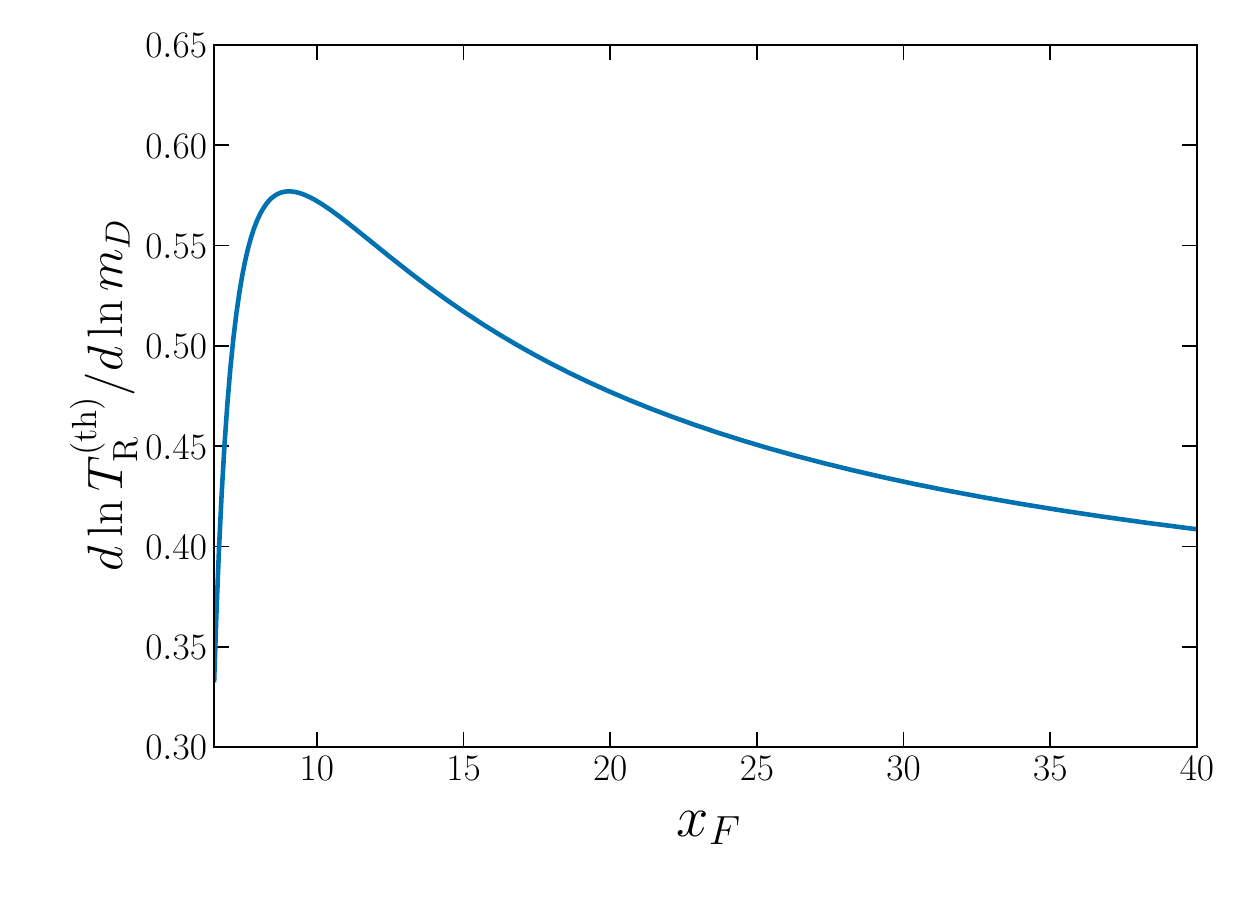}
    \caption{
    The effective power-law exponent
    $d\ln T_{\mathrm{R}}^{(\rm{th})}/d\ln m_D$
    along a contour of fixed relic abundance,
    shown as a function of $x_F=m_D/T_F$.
    }
    \label{fig:TRH}
\end{figure}

\section{Dark Matter from Right-Handed Neutrino Decays}
\label{app: dark matter emission}
In this Appendix, we discuss the irreducible production of dark matter associated with right-handed neutrino decays.
\subsection{Setup and Two-body Decay Rate}
For simplicity, 
we consider $\mathrm{SU}(2)_L\times \mathrm{U}(1)_Y$ theory 
in a symmetric phase
with
one lepton doublet $\ell=(\nu,e)^T$ in the left-handed Weyl fermion, a Higgs doublet
$H=(H^+,H^0)^T$, a Majorana singlet left-handed Weyl fermion $\bar{N}_R$ of mass $M_N$,
and a pseudo-Dirac doublet dark matter.
As we work in the $\mathrm{SU}(2)_L\times \mathrm{U}(1)_Y$ symmetric phase,
the leptons, the Higgs doublet and gauge bosons are massless.
In the following discussion, the mass splittings within the dark matter multiplet are irrelevant, and we therefore treat it as a vector-like Dirac doublet,
\begin{align}
\psi^0=\mqty(\chi^0\\\bar{\chi}^{0\dagger})\ , \quad
\psi^+ = \mqty(\chi^+\\\bar{\chi}^{-\dagger})\ ,
\end{align}
where $\psi^0$ and $\psi^+$ form a doublet of $\mathrm{SU}(2)_L$.

The Yukawa 
interaction 
and Majorana mass term are
\begin{align}
 \mathcal L\supset
 -Y \bar{N}_R\,\epsilon_{ab}\ell^aH^b
    -\frac{M_N}{2}\bar{N}_R \bar{N}_R+\mathrm{h.c.},
 \qquad \epsilon_{12}=+1.
\end{align}
Thus $\epsilon_{ab}\ell^aH^b=\nu H^0-eH^+$.
The  $\mathrm{SU}(2)$ doublet $\chi$ and 
its mass partner
$\bar{\chi}$ 
have a gauge-invariant Dirac mass and
interact 
with this sector only through gauge interactions.

The decay rate of the right-handed neutrino is given by,
\begin{align}\label{eq:gamma2}
 \Gamma_N=
\Gamma(\bar{N}_R\to \ell +H) + 
\Gamma(\bar{N}_R\to \ell^\dagger +H^\dagger)
=\frac{|Y|^2M_N}{8\pi}\ .
\end{align}
The decay rate for each individual component of the doublet, with particles and antiparticles treated separately, is one quarter of this value.

\subsection{Associated Production and Channel Counting}

The four-body process proceeds through
\begin{align}
\bar{N}_R\to\ell+ H+ V^*\ ,\qquad V^*\to \psi\bar{\psi}
  \ ,
 \qquad V=B,W^a\ .
\end{align}
Emission from both the lepton and Higgs lines is included.

There are six channels for the 
associated dark matter production,
\begin{align}
\begin{gathered}
 \nu H^0\psi^0\bar \psi^0\ ,\quad
 \nu H^0\psi^+\bar\psi^+\ ,\quad
 e^-H^+\psi^0\bar \psi^0\ ,\quad
 e^-H^+\psi^+\bar{\psi}^+\ ,\\
 \nu H^+\psi^0\bar{\psi}^+\ ,\qquad
 e^-H^0\psi^+\bar \psi^0\ .
\end{gathered}
\end{align}
Their charge conjugates have equal rates at tree level in
this single-Yukawa model. Accordingly, 
the total
initial spin averaged and the final spin summed
squared amplitude is given by
\begin{align}
 S_4= 2\sum_{c=1}^{6}\frac12\sum_{\mathrm{spins}} \qty|\mathcal{M}|^2 \ .
 \label{eq:S4}
\end{align}
Here, $\mathcal{M}$ 
denotes the amplitude for each channel,
which include the dark matter production from both the lepton and Higgs lines.
Notice that, after summing over all doublet components in the final
state, the interference terms proportional to $g^2g^{\prime 2}$
vanish because they are proportional to $\Tr(Y_D T^a)=0$.
In our numerical analysis, we made use of 
FeynArts~\cite{Hahn:2000kx} and FeynCalc~\cite{Mertig:1990an,Shtabovenko:2016sxi,
Shtabovenko:2020gxv,Shtabovenko:2023idz}.

\subsection{Four-body Phase Space}

We adopt the phase-space convention
\begin{align}
d\Phi_4=(2\pi)^4\delta^{(4)}
 \left(P-\sum_i p_i\right)
\prod_{i=1}^{4}\frac{d^3p_i}{(2\pi)^3\,2E_i}\ ,
\end{align}
with which the decay rate is given by,
\begin{align}
\Gamma_4=\frac{1}{2M_N}\int d\Phi_4\,S_4  \ .
\end{align}
For later purpose, we introduce kinematica variables,
\begin{align}
Q=p_\ell+p_H,\quad K=p_\psi+p_{\bar \psi}\ ,\quad x=\frac{Q^2}{M_N^2}\ ,\quad z=\frac{K^2}{M_N^2}\ ,
 \quad r=\frac{m_D}{M_N}\ .
\end{align}
With these parameters, the recursive factorization leads to
\begin{align}
 d\Phi_4=
 \frac{d Q^2\,d K^2}{(2\pi)^2}
 d\Phi_2(P;Q,K) d\Phi_2(Q;p_\ell,p_H) d\Phi_2(K;p_\psi,p_{\bar \psi})\ .
\end{align}

Choose $Q$ along $+z$ in the $\bar N_R$ rest frame, and then,
$K$ points along $-z$.
In the $Q$ and $K$ rest frames, choose parallel spatial axes
and define
\begin{align}
c_\ell=\cos\theta_\ell\ ,\qquad
c_\psi=\cos\theta_\psi\ ,
\end{align}
respectively.
Both polar angles are measured from the common
$+z$ direction.
Define the relative azimuth by
\begin{align}
 \phi=\varphi_\psi-\varphi_\ell \pmod{2\pi}\ .
\end{align}
Boosts along the $z$ axis leave the transverse momentum
components unchanged, so this relative azimuth is the same
in the parent rest frame.

The variables $x$ and $z$ determine the invariant masses
of the two pairs, $\sqrt{Q^2}=M_N\sqrt{x}$ and
$\sqrt{K^2}=M_N\sqrt{z}$.
Then, choosing the common azimuth such that
$\varphi_\ell=0$ and $\varphi_\psi=\phi$, the
three-momenta are
\begin{gather}
 \boldsymbol p_\ell^*
=\frac{M_N\sqrt{x}}{2}
\left(s_\ell,\,0,\,c_\ell\right)\ ,
\qquad
 \boldsymbol p_H^*=-\boldsymbol p_\ell^*\ ,\\
 \boldsymbol p_\psi^*
=\frac{M_N\sqrt{z-4r^2}}{2}
 \left(
 s_\psi c_\phi,\,
s_\psi s_\phi,\,
 c_\psi
 \right)\ ,
 \qquad  \boldsymbol p_{\bar\psi}^*=-\boldsymbol p_\psi^*\ ,
\end{gather}
where $s_x=\sin \theta_x$.
Here, the stars denote quantities evaluated in the
corresponding pair rest frame.
Thus, $(x,z)$ fix the momentum magnitudes, while
$(c_\ell,c_\psi,\phi)$ specify their directions
up to an overall spatial rotation.

In the parent rest frame, the pair energies and
common momentum magnitude are
\begin{align}
 E_Q=\frac{M_N}{2}(1+x-z)\ ,\qquad 
 E_K=\frac{M_N}{2}(1-x+z)\ , \qquad
 |\boldsymbol Q|=|\boldsymbol K|
 =\frac{M_N}{2}\sqrt{\lambda(1,x,z)}\ ,
\end{align}
where
\begin{align}
 \lambda(a,b,c)=a^2+b^2+c^2-2ab-2ac-2bc\ ,
\end{align}
which corresponds to the kinematical variables for the two body decay of $\bar{N}_R$ into particles with squared masses $Q^2$ and $K^2$.

At fixed pair invariant masses, the delta functions in each
two-body phase space fix the daughter momentum magnitude,
leaving only the solid angle,
\begin{align}
d\Phi_2(X;p_1,p_2)
=\frac{|\boldsymbol p_*|}{16\pi^2\sqrt{s}}\,d\Omega
=\frac{\sqrt{\lambda(s,m_1^2,m_2^2)}}{32\pi^2s}\,d\Omega\ ,
 \qquad s=X^2\ .
\end{align}
Here, $\boldsymbol p_*$ is evaluated in the rest frame of $X$.

For an unpolarized decay, integrating over the direction
of $\boldsymbol Q$ and the common azimuth yields
\begin{align}\label{eq:phase}
 d\Phi_4=
 \frac{M_N^4}{2^{14}\pi^6}\,
 \rho(x,z)\,
 dx\,dz\,dc_\ell\,dc_\psi\,d\phi\ ,
\end{align}
where
\begin{align}
 \rho(x,z)=
 \sqrt{\lambda(1,x,z)}
 \sqrt{1-\frac{4r^2}{z}}\ .
\end{align}
The integration domain is
\begin{align}
\begin{gathered}
 4r^2\le z\le1\ ,\qquad
 0\le x\le(1-\sqrt z)^2\ ,\\
 -1\le c_\ell,c_\psi\le1,\qquad
 0\le\phi<2\pi\ .
\end{gathered}
\end{align}
The rate vanishes for $2r\ge1$.
As a normalization check, the fully massless
phase-space volume is
\begin{align}
\int d\Phi_4\big|_{r=0}
 =\frac{M_N^4}{24576\pi^5}\ .
\end{align}
The decay rate is then given by,
\begin{align}
    \Gamma_4 = \frac{M_N^3}{2^{15}\pi^6}\int  \rho(x,z)\,S_4\,
 dx\,dz\,dc_\ell\,dc_\psi\,d\phi\ ,
\end{align}
where $S_4$ is given in Eq.\,\eqref{eq:S4}.
\subsection{Soft Emission Limit}
The full four-body phase-space calculation in the preceding subsection
requires numerical integration. In the hierarchical limit $m_D \ll M_N$, however, the leading logarithmic contribution admits a simple
analytic expression, which we derive below.

We begin with the exact recursive factorization of the four-body
phase space,
\begin{equation}
 d\Phi_4
 =\frac{d s}{2\pi}\,
d\Phi_3(P;p_\ell,p_H,q)\,
d\Phi_2(q;p_\psi,p_{\bar{\psi}})\ .
 \label{eq:phase-factor}
\end{equation}
Here, $q=p_{\psi}+p_{\bar\psi}$ and $s=q^2$.
At each fixed $s$, the three-body phase space treats $q$ as the momentum
of a particle of mass $\sqrt{s}$. The integration range is
$4m_D^2\le s\le M_N^2$.
For $m_D^2 \ll M_N^2$, logarithmically enhanced contributions arise from
$s \ll M_N^2$. The highest logarithmic power is generated when the emitted pair
is also soft in energy and nearly collinear to either hard final-state
particle. We therefore focus on this region.

The dark matter pair is produced by the splitting of an off-shell gauge
boson emitted from a final-state lepton or Higgs line. The four-body
amplitude can accordingly be written as
\begin{equation}
 \mathcal M_{4}
 =\mathcal M_3^\mu\,
 \frac{-g_{\mu\nu}}{s}\,
 g' Y_{D}J_{D}^\nu
  \ ,
 \qquad
 \label{eq:amp-factor}
\end{equation}
where $J_D^\nu =\bar{u}(p_\psi)\gamma^\nu v(p_{\bar{\psi}})$.
We first consider the $\mathrm{U}(1)_Y$ contribution, where
$Y_D=1/2$ denotes the $\mathrm{U}(1)_Y$ hypercharge of the dark matter.
The three-body amplitude,
\begin{align}
\mathcal{M}_3^\mu:\quad
\bar{N}_R \to \ell + H + B^{\mu *}(q)\ ,
\end{align}
includes emission from both the lepton and Higgs lines. For the hierarchy
$m_D\ll M_N$, its leading soft contribution takes the eikonal
form~\cite{Catani:1999ss},
\begin{align}
     \mathcal M_3^\mu\simeq g'\mathcal M_0j^\mu\ ,
 \qquad j^\mu=\left(\frac{Y_H p_H^\mu}{p_H\cdot q}
+\frac{Y_\ell p_\ell^\mu}{p_\ell\cdot q}\right)\ ,
 \qquad Y_H=-Y_\ell=\frac12\ ,
\end{align}
where $\mathcal{M}_0$ is the Born amplitude for $\bar{N}_R \to \ell + H$.
The leading soft current depends only on the charges and momenta of the
emitting particles and is therefore spin independent.

At fixed $s=q^2$, summing over the pair spins and integrating over
the two-body phase space gives
\begin{align}
 \int d\Phi_2 \sum_{\rm spins} J_D^\mu J_D^{\nu *}
 &=
 \frac{s}{6\pi}
 \sqrt{1-\frac{4m_D^2}{s}}
 \left(1+\frac{2m_D^2}{s}\right)
 \left(-g^{\mu\nu}+\frac{q^\mu q^\nu}{s}\right)
 \nonumber\\
 &\equiv
\frac{s}{6\pi} R_{D}(s)
 \left(-g^{\mu\nu}+\frac{q^\mu q^\nu}{s}\right)\ ,
\end{align}
where the sum runs over the spins of the final-state dark matter particles.

Neglecting recoil at leading order in the soft expansion, the three-body
phase space factorizes as
\begin{equation}
 d\Phi_3(P;p_\ell,p_H,q)
 \simeq
 \frac{d^3\mathbf{q}}{(2\pi)^3\,2\omega}\,
 d\Phi_2(P;p_\ell,p_H)\ .
 \label{eq:phase-recoil}
\end{equation}
Combining this with the factorized amplitude,
\begin{align}
 \mathcal{M}_3^\mu \mathcal{M}_3^{\nu *}
 =
 |\mathcal{M}_0|^2\,
 g^{\prime\,2}\,
 j^\mu j^{\nu *}\ ,
\end{align}
we obtain the four-body decay rate normalized to the Born rate,
\begin{align}
 \frac{\Gamma_4}{\Gamma_N}
 \simeq
 2g^{\prime\,4}Y_D^2
 \int \frac{d s}{2\pi}\,
 \frac{d^3\mathbf{q}}{(2\pi)^3\,2\omega}\,
 \left(-j^\mu j_\mu^*\right)
 \frac{R_{D}(s)}{6\pi s}\ .
\end{align}
Here, we have used $q^\mu j_\mu=0$. The prefactor of two counts the
two degenerate components of the dark matter doublet.

In the $\bar{N}_R$ rest frame, we evaluate the soft current at the Born
momenta and parameterize the pair momentum as
\begin{align}
    p_\ell = \frac{M_N}{2}(1,0,0,1)\ , \quad \mathbf{p}_H=-\mathbf{p}_\ell\ , \quad 
    q = \omega(1,\beta_qs_q,0,\beta_q c_q)\ , \quad \beta_q = \sqrt{1-\frac{s}{\omega^2}} \ .
\end{align}
The squared soft current then becomes
\begin{align}
    -j^\mu j_\mu^*= \frac{-2Y_HY_\ell p_H\cdot p_\ell}{(p_H\cdot q)(p_\ell \cdot q)} = \frac{1}{\omega^2(1-\beta_q^2 c_q^2)}\ .
\end{align}
After integrating over the azimuthal angle, the normalized rate reduces to
\begin{align}
 \frac{\Gamma_4}{\Gamma_N}
 \simeq
 \frac{g'^4}{192\pi^4}
 \int \frac{d s}{s}\,
\int \frac{d\omega}{\omega}
\int dc_q \frac{\beta_q R_D(s)}{1-\beta_q^2 c_q^2}\   .
\end{align}

The three logarithmic enhancements originate from the integrations over
the pair invariant mass squared $s$, the angular variable $c_q$, and the pair energy
$\omega$. Writing $q^0=\omega$ and $c_q=\cos\theta_q$, the collinear regions correspond to $c_q\simeq\pm1$, or equivalently $\theta_q\simeq0,\pi$. Near the lepton-collinear direction, the
logarithmically enhanced region is characterized by
\begin{align}
\int \frac{ds}{s}
\int \frac{d\theta_q^2}{\theta_q^2}
\int \frac{d\omega}{\omega}\ ,
\qquad
4m_D^2 \ll s \ll \omega^2 \ll M_N^2\ ,
\qquad
\frac{s}{\omega^2}\ll \theta_q^2\ll1\ .
\end{align}
An analogous region occurs around the Higgs direction. In these regions,
we may set $\beta_q\to1$ and $R_D(s)\to1$ in the numerator while retaining the
virtuality in the collinear denominators. To leading logarithmic accuracy,
the rate therefore becomes
\begin{align}
\frac{\Gamma_4}{\Gamma_N}
\simeq
\frac{g'^4}{192\pi^4}
\int_{2m_D}^{\Lambda}\frac{d\omega}{\omega}
\int_{4m_D^2}^{\omega^2}\frac{ds}{s}
\int_{-1}^{1}dc_q\,
\frac{1}{1-c_q^2+(s/\omega^2)c_q^2}\ .
\end{align}
Here, $\Lambda$ is a soft cutoff satisfying
$2m_D\ll\Lambda\ll M_N$. We retain the $s$-dependent term in the denominator because it regulates the
collinear singularities at $c_q=\pm1$.

Extracting the leading logarithm yields
\begin{align}
\left.\frac{\Gamma_4}{\Gamma_N}\right|_{\mathrm{LL}}
\sim
\frac{g^{\prime 4}}{12\times192\pi^4}
\qty(\log\frac{M_N^2}{m_D^2})^3\ ,
\qquad
m_D\ll M_N\ .
\end{align}
The dependence on the intermediate scale $\Lambda$ affects only subleading logarithmic terms if $\Lambda$ is held at a fixed small fraction of the parent-neutrino mass as the
hierarchy increases. In particular, it must satisfy
\begin{align}
\log\frac{M_N}{m_D}\gg\log\frac{M_N}{\Lambda}\ .
\end{align}
The same soft kinematics apply to emission of off-shell weak-isospin
$W$ bosons. In the unbroken electroweak approximation, summing over the
three weak gauge bosons and all degenerate doublet components gives
\begin{align}
\left.\frac{\Gamma_4}{\Gamma_N}\right|_{\mathrm{LL}}
\sim
\frac{g^{\prime 4}+3g^4}{12\times192\pi^4}
\qty(\log\frac{M_N^2}{m_D^2})^3\ ,
\qquad
m_D\ll M_N\ .
\label{eq:leading log}
\end{align}

For a sufficiently large hierarchy $m_D\ll M_N$, the fixed-order
estimate of the $\psi\bar \psi$ emission fraction can become of order unity
even though the gauge couplings remain perturbative. This signals a
breakdown of the fixed-order expansion. At relative order $g^4$, the
interference between the Born amplitude and the two-loop amplitude
containing a virtual $\psi\bar \psi$ loop contributes a negative leading
triple logarithm to the rate without a $\psi\bar \psi$ pair. 
When these corrections become large, a consistent prediction requires
resummation of real and virtual radiation, including multiple emissions~\cite{Hoang:1993ns}.
Our fixed-order treatment of pair emission
therefore requires
\begin{equation}
 \frac{\Gamma_{4,\mathrm{LL}}^{\mathrm{EW}}}{\Gamma_N}
 =
 \frac{\alpha_Y^2+3\alpha_2^2}{144\pi^2}
 \qty(\log\frac{M_N^2}{m_D^2})^3
 \ll 1\ ,
 \qquad
 \alpha_Y=\frac{g'^2}{4\pi}\ ,\quad
 \alpha_2=\frac{g^2}{4\pi}\ .
\end{equation}
Here the numerator and denominator include the corresponding sums over
weak-doublet components and charge-conjugate channels. This is a necessary
condition for treating pair emission as a small correction; other enhanced
electroweak corrections relevant to the observable must also remain under
control.

\bibliographystyle{apsrev4-1}
\bibliography{main}

\end{document}